\documentclass[conference]{IEEEtran}
\IEEEoverridecommandlockouts
\usepackage{cite}
\usepackage{amsmath,amssymb,amsfonts,amsthm}
\usepackage{algorithmic}
\usepackage[lined, ruled, linesnumbered, vlined]{algorithm2e}
\usepackage{graphicx}
\usepackage{subcaption}
\usepackage{textcomp}
\usepackage{xcolor}
\usepackage{bm}
\usepackage{authblk}
\usepackage[hidelinks]{hyperref}
\def\BibTeX{{\rm B\kern-.05em{\sc i\kern-.025em b}\kern-.08em
    T\kern-.1667em\lower.7ex\hbox{E}\kern-.125emX}}
\usepackage[hang,flushmargin]{footmisc}

\newtheorem{definition}{\bf Definition}	
\newtheorem{theorem}{\bf Theorem}		
\newtheorem{lemma}{\bf Lemma}		
\newcounter{appdx}
\newcommand{\s}         {s}

\newcommand{\red}[1]{{\color{black} #1}} 
\newcommand{\blue}[1]{{\color{black} #1}}
\newcommand{\orange}[1]{{\color{black} #1}} 
\newcommand{\dgreen}[1]{{\color{black} #1}} 
\newcommand{\purple}[1]{{\color{black} #1}} 
\newcommand{\gray}[1]{{\color{black} #1}}
\newcommand{\revision}[1]{{\color{black} #1}} 

\definecolor{dgreen}{rgb}{0,0.655,0.149}

\newcommand{\black}[1]{{\color{black} #1}}

\makeatletter
\newcommand{\nosemic}{\renewcommand{\@endalgocfline}{\relax}}
\newcommand{\dosemic}{\renewcommand{\@endalgocfline}{\algocf@endline}}
\let\oldnl\nl
\newcommand{\nonl}{\renewcommand{\nl}{\let\nl\oldnl}}
\makeatother

\SetKwInOut{Init}{Initialize}

\IEEEaftertitletext{\vspace{-0.35in}}
\makeatletter
\renewenvironment{proof}[1][\proofname]{\par
  \pushQED{\qed}%
  \normalfont
  \topsep0pt \partopsep0pt 
  \trivlist
  \item[\hskip\labelsep
        \itshape
    #1\@addpunct{.}]\ignorespaces
}{%
  \popQED\endtrivlist\@endpefalse
  \addvspace{6pt plus 6pt} 
}
\makeatother

\IEEEoverridecommandlockouts\IEEEpubid{\makebox[\columnwidth]{ 979-8-3503-0322-3/23/\$31.00 $\copyright$2024 IEEE \hfill}\hspace{\columnsep}\makebox[\columnwidth]{ }}
\begin{document}

\title{Infiltrating the Sky: Data Delay and  Overflow Attacks in Earth Observation Constellations}

\vspace{-2mm}
\author[*]{\rm Xiaojian~Wang}
\author[*]{\rm Ruozhou~Yu}
\author[$\dag$]{\rm Dejun~Yang}
\author[$\ddag$]{\rm Guoliang~Xue\vspace{-3.5mm}}
\affil[ ]{$^*$North Carolina State University, $^{\dag}$Colorado School of Mines, $^{\ddag}$Arizona State University\vspace{3mm}}

\maketitle


\begin{abstract}
Low Earth Orbit (LEO) Earth Observation (EO) satellites have changed the way we monitor Earth. Acting like moving cameras, EO satellites are formed in constellations with different missions and priorities, and capture vast data that needs to be transmitted to the ground for processing. However, EO satellites have very limited downlink communication capability, limited by transmission bandwidth, number and location of ground stations, and small transmission windows due to high-velocity satellite movement. To optimize resource utilization, EO constellations are expected to share communication spectrum and ground stations for maximum communication efficiency.

In this paper, we investigate a new attack surface exposed by resource competition in EO constellations, targeting the delay or drop of Earth monitoring data using legitimate EO services. Specifically, an attacker can inject high-priority requests to temporarily preempt low-priority data transmission windows. Furthermore, we show that by utilizing predictable satellite dynamics, an attacker can intelligently target \emph{critical data} from low-priority satellites, either delaying its delivery or irreversibly dropping the data. We formulate two attacks, the data delay attack and the data overflow attack, design algorithms to assist attackers in devising attack strategies, and analyze their feasibility or optimality in typical scenarios. We then conduct trace-driven simulations using real-world satellite images and orbit data to evaluate the success probability of launching these attacks under realistic satellite communication settings. We also discuss possible defenses against these attacks.
\end{abstract}




\section{Introduction}

\noindent Low Earth Orbit (LEO) Earth observation (EO) constellations have revolutionized Earth monitoring by employing numerous low-orbit, low-cost small satellites, rather than a few high-cost, high-altitude large satellites targeted only at specific regions.
EO constellations enable continuous, high-resolution imaging of the entire Earth's surface, significantly reducing revisit times and enhancing observation frequency.
EO data is widely used in various fields, such as agriculture, forestry, urban planning, and disaster management~\cite{le2020space}.

As EO satellites continuously scan Earth's surface, they generate vast monitoring data that must be transmitted to the ground for processing and analysis.
However, several factors hinder efficient satellite data downlinking.
Firstly, the geographical requirements and high costs of constructing ground stations limit their number and distribution, resulting in few available windows for downlinking data~\cite{Planet-communication-network}.
Secondly, the high velocity of satellites results in short contact windows with ground stations, typically lasting only 7-10 minutes~\cite{devaraj2017dove,tao2023transmitting}.
Third,
due to the physical limitations of radio frequency spectrum allocation and atmospheric absorption, data rates from satellite to ground station remain limited to an average of 160 Mbit/s~\cite{Planet_downlink_rate}, even with a variety of bands and advanced antenna design or modulation and coding schemes.
These factors lead to insufficient downlink resources for the satellite data, necessitating onboard storage and resulting in delayed transmission, ranging from a few hours to several days~\cite{course-intro-planet}.

A trend in EO involves multiple types of constellations collaboratively achieving mission goals.
\dgreen{For instance, a low-priority EO constellation consists of satellites with low-resolution cameras sweeping the Earth's surface for continuous monitoring, while a much smaller high-priority constellation with high-resolution cameras focus on specific Areas of Interest (AoIs) based on results from low-priority satellites and/or requests from users~\cite{course-intro-planet}.}
\revision{Furthermore, constellations are increasingly supporting user-scheduled time-sensitive sensing tasks to enable critical use cases such as rapid disaster response or real-time emergency management~\cite{assured_tasking}, and will prioritize transmission of such user-scheduled high-priority tasks over the default (low-priority) monitoring tasks via fast tracked data delivery to the ground~\cite{Fast_Track_Delivery}.}

\dgreen{In face of limited ground stations, the low- and high-priority constellations}
may share communication channels and ground stations to maximize the chance and data rate of downlink communication~\cite{colton2020merging}.
When there is a high-priority request such as monitoring the spread of forest fires at a specific location from a user, the low-priority satellites may be preempted of their communication resources to prioritize high-value data transmission, in which case the low-priority monitoring data must be stored until there is a future transmission opportunity.

This paper investigates the possibility that an attacker may utilize the sharing of communication resources and opportunities among mixed-priority EO satellites to launch intelligent attacks targeting critical data captured by low-priority EO satellites.
Such attacks include, for instance, delaying the downlink of target data captured at a certain time and location, or dropping such data from the internal storage of a satellite before it can be downlinked.
In doing so, for instance, the attacking entity (potentially a nation or organization) may aim to prevent downlink and analysis of sensitive information, such as regional warfare strategies or illegal operations, without launching large-scale, easily detectable attacks such as disabling a large number of ground stations (as happened at the beginning of the Russia-Ukraine conflict)~\cite{attack_ground_station} or jamming all satellite communication channels~\cite{hofmann2018satellite, taricco2022jamming}.

We explore a new attack surface that exploits the competition for limited downlink resources between different types of EO satellites.
Specifically, when an attacker deliberately schedules tasks from high-priority satellites that shares the limited downlink resources with a low-priority satellite, the latter will have to wait for the next transmission opportunity, leading to data on the low-priority satellite being delayed.
If data accumulated on the low-priority satellite exceeds the internal storage limit, some data will be dropped.
\dgreen{Specifically, by leveraging the predictable dynamics of satellites, including orbit information, scheduling policies, and queue status on the satellite, an attacker can intelligently select its attack strategy, such as how many/which transmission windows to attack, 
to make its attack more viable and less detectable.}
In this paper, we formalize data delay and data overflow problems, where an attacker targets a specific piece of data to delay or drop, and design two algorithms to solve these problems.
Our attack can be planned or even launched before the target data is captured by a satellite, and utilize only legitimate services provided by different EO constellations, thus making the attack easy to launch and the attack target hard to locate or protect.

The contribution of this paper is listed as follows:
\begin{itemize}

    \item We explore and propose a new type of attack targeting data captured by certain EO satellites that share communication resources with others, utilizing the shared communication resources to delay or prevent the downlink of the target data    via legitimate EO service requests.
    \item We formulate two possible attack goals of an adversary: the \emph{data delay} attack and the \emph{data overflow} attack, and devise algorithms to find feasible attack strategies considering orbital dynamics, bandwidth, storage, and attack costs. We analyze the feasibility or optimality of our algorithms in typical attack scenarios.
    \item We evaluate the attacks using real-world satellite orbital dynamics and imaging data, demonstrating their practical implications and effectiveness.
\end{itemize}

\noindent
\textbf{Organization.}
\S\ref{sec:related-work} introduces background and related work.
\S\ref{sec:system-model} formalizes the system model.
\S\ref{sec:overview} describes the overview of two attacks.
\S\ref{sec:delay-attack} and \S\ref{sec:buffer-overflow-attack} present the data delay attack and the data overflow attack.
\S\ref{sec:practical-consideration} addresses practical considerations of launching the attacks.
\S\ref{sec:eval} presents the evaluation results.
\S\ref{sec:discussion} describes some countermeasures against the proposed attacks.
\S\ref{sec:conclusion} concludes the paper.
\S \red{Appendix provides proofs of lemmas and theorems of this paper.}



\section{Background and Related Work} 
\label{sec:related-work}
\subsection{Earth Observation Constellations}
\noindent 
Satellites positioned at altitudes below 2000km are called LEO satellite.
When used for Earth observation, these satellites are formed in EO constellations that continuously monitor the Earth's surface with high resolution and low costs, and can offer detailed and frequent data for tracking environmental changes, managing disasters, monitoring resources, etc.

\begin{figure}
    \centering\includegraphics[width=0.45\textwidth]{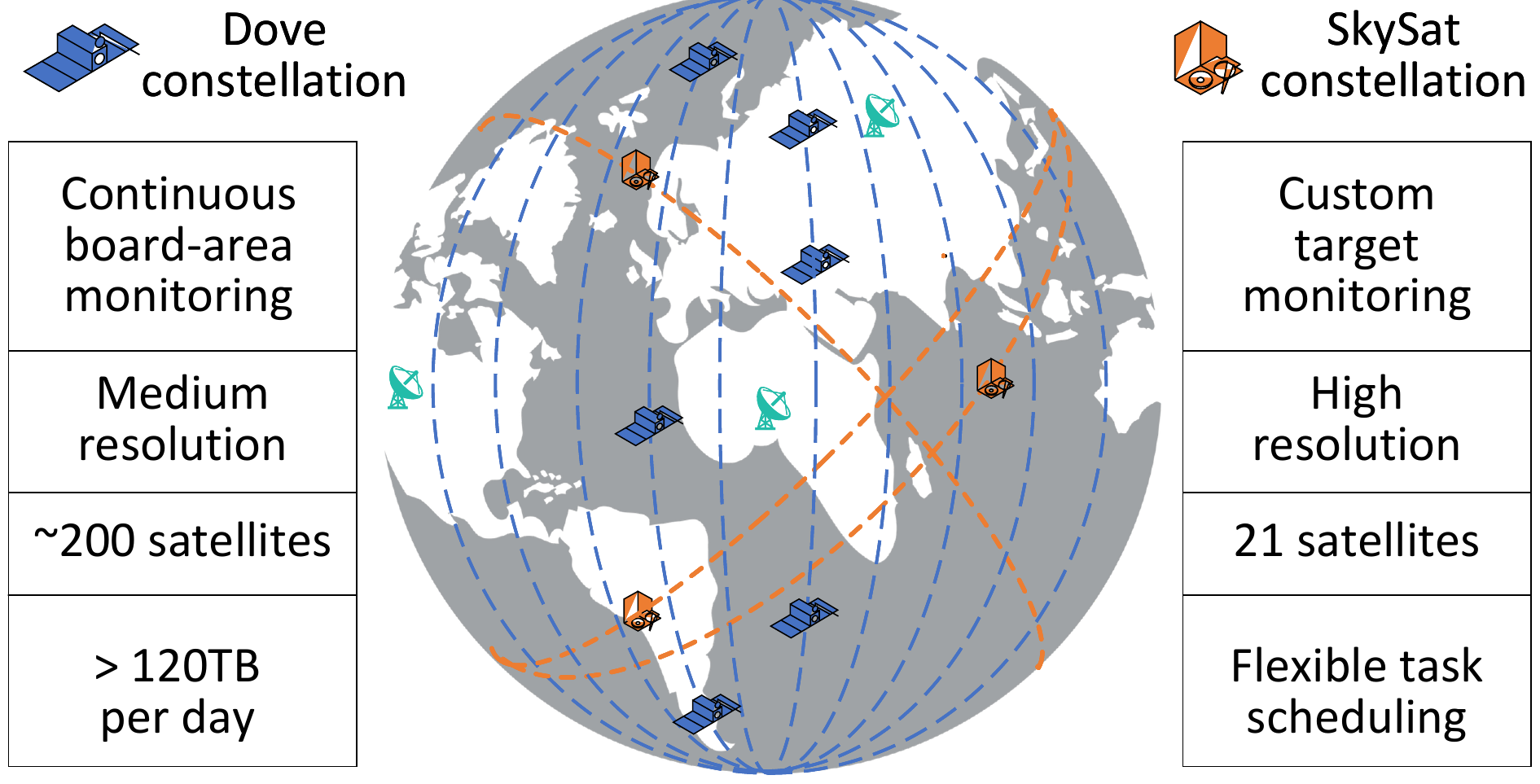}
    \vspace{4mm}
    \caption{EO constellations operated by Planet Labs~\cite{course-intro-planet}.}
    \label{fig:constellation}
\end{figure}

Here, we use the constellations operated by Planet Labs~\cite{course-intro-planet} as an example to illustrate the structure of EO constellations\footnotemark{}.
\footnotetext{\blue{While our attacks are motivated by Planet Labs, they can be applied to any constellations that (partially) share ground stations for data downlink.}}%
Planet Labs hosts the world's largest commercial EO constellations, with over 150 satellites currently in LEO orbits~\cite{colton2020merging}.
As shown in Fig.~\ref{fig:constellation}, Planet Labs currently has two constellations: 
the Dove constellation and the SkySat constellation.
The Dove constellation offers continuous imaging with daily medium-resolution imagery of 3.7 meters per pixel, covering the entire Earth's surface.
The Dove constellation has a daily data collection capability of 200 million km$^2$~\cite{Planet-communication-network} and captures 120TB of data per day~\cite{tao2023transmitting}.
The SkySat constellation, comprising 21 satellites, focuses on high-resolution imagery of areas of interest at 0.65 meters per pixel, and offers on-demand collection with less than 15\% cloud cover, prioritizing areas with high demand.
This constellation allows users to submit tasks with flexible scheduling options, including pre-scheduled, monthly, weekly, daily, or sub-daily for specified areas and time frames, featuring different delivery strategies for various orders~\cite{course-intro-planet, Planetlab_tasking}.
Specifically, the two constellations can work collaboratively for an EO user.
For instance, a user may first use the large amount of data generated by Dove satellites to locate areas of interest, and then use the SkySat satellites to capture high-quality data of those areas at specific times.
Currently, there are no Inter-Satellite Links (ISLs) implemented for the constellations~\cite{Planet-communication-network}.

Ground stations of the constellations utilize S-band for uplink and X-band for downlink communications~\cite{Planet-Labs-Ground-Station-Network-2016}.
\black{Currently, there are only a handful of ground stations globally to support downlink transmissions~\cite{48-ground-stations, tao2023transmitting}.}
\black{A typical pass, during which a downlink X-band lock is established and maintained with a ground station for reliable communication and data transfer, lasts only 7-10 minutes and requires precise antenna pointing with errors of less than 0.2 degrees~\cite{foster2015orbit}.}
\black{The data rates from satellite to ground station are limited to an average of 160 Mbit/s~\cite{Planet_downlink_rate}.}
Given the limited communication resources and windows, it is expected that most of a satellite's captured data needs to be stored in its internal storage for some period of time before being downlinked via a ground station.
This has motivated the study of buffer management in these satellites.

\subsection{Related Work}

\noindent
Our attacks are related to existing works that study buffer management for satellites/spacecrafts given the limited communication capabilities.
While many works have studied scheduling data collection, downlink or in-orbit computing for improving throughput and reducing latency and data loss~\cite{tao2023transmitting, vasisht2021l2d2,taoknown,herrmann2022autonomous, rabideau2016managing, knight2014leveraging, rabideau2017managing, lin2005daily,lyu2023falcon}, their scope is limited to a single satellite or constellation, and neglects the inter-play among satellites between different constellations.
Beyond buffers in the space, queuing analysis has been widely studied in traditional and wireless sensor networks.
These studies focus on evaluating the performance of network systems by examining the delay and throughput characteristics of data packets~\cite{kesselman2001buffer, xu2014analytical, gao2022buffer, song2024inversion, tan2021regularization, tan2019joint,huang2022traffic, gao2022gearbox}.
Some works have targeted the utilization of queue dynamics for buffer overflow distribution analysis~\cite{kempa2020distribution, kempa2020time}, but they are limited to specific queue models with certain input and output patterns.
\blue{Some works have examined buffer attacks on single-node High Performance Computing (HPC) systems or targeted the memory bus of a single machine~\cite{zhang2019tail,hunger2015understanding}, which address a inherently different problem space from ours.}
To our knowledge, no existing work has studied buffer management through an adversarial lens, where an adversary can utilize the knowledge towards the scheduling policy to launch data-oriented attacks.



\section{System Model}
\label{sec:system-model}

\subsection{EO Constellations in Low Earth Orbit}
\label{subsec:constellation_model}

\begin{figure}[t]
    \centering\includegraphics[width=0.45\textwidth]{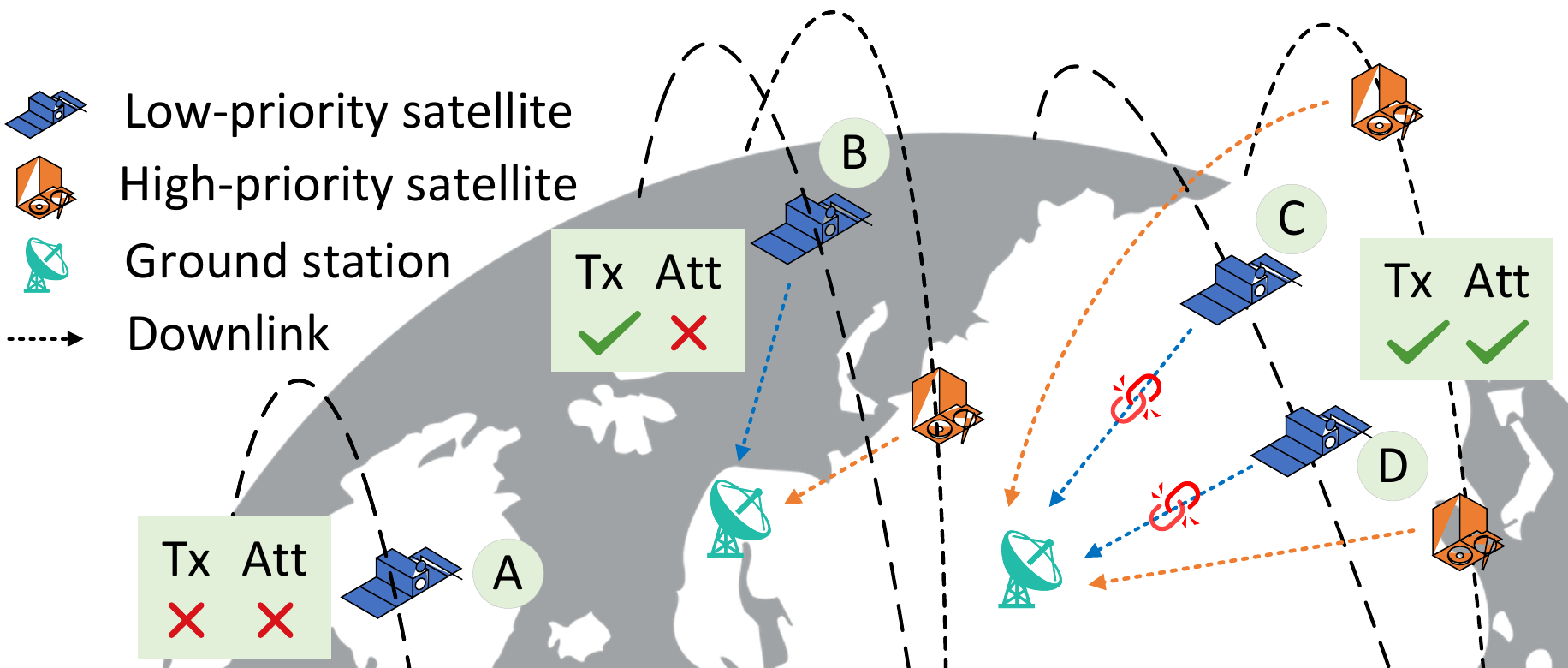}
    \vspace{4mm}
    \caption{EO constellations example consists of 4 low-priority satellites, 3 high-priority satellites, \purple{and 2 ground stations with 2  antennas each}.
     \textsf{Tx} represents the transmissible indicator, and \textsf{Att} represents the attackable indicator. 
    }
    \label{fig:attack_model}
\end{figure}

\noindent As shown in Fig.~\ref{fig:attack_model}, we consider two EO constellations of LEO satellites: a high-priority constellation $S_h$ and a low-priority constellation $S_l$. 
They use the same set of ground stations with a limited number of antennas, and compete for limited downlink resources over the same spectrum bands.
\purple{Data collected by high-priority satellites has a higher priority and needs to be transmitted to ground stations as quickly as possible.}
Data is collected on a per data unit (e.g., per image) basis. 
The system time is discretized into time slots $\mathbb{T} = \{0, 1, 2, \ldots, \bm{T}\}$.
\purple{The data amount collected from $s_i \!\in\! S_l$ at $t \!\in\! \mathbb{T}$ is denoted as  $I_{s_i}(t)$.}

\begin{figure}
    \centering\includegraphics[width=0.37\textwidth]{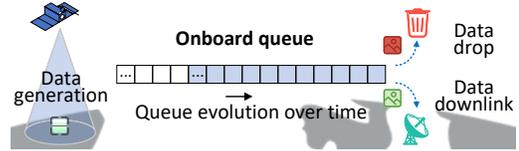}
    \vspace{4mm}
    \caption{Onboard queue evolution. 
    Low-priority satellites scan the Earth's surface, generating onboard queue input data.
    Data in the queue is either downlinked to a ground station or dropped due to limited storage capacity.}
    \label{fig:queue-model}
\end{figure}

Each satellite $s_i \in S_l$ has a limited internal storage with capacity $c_{s_i}$.
\purple{As shown in Fig.~\ref{fig:queue-model}, captured Earth data is stored in the storage before being transmitted to a ground station. }
We define the \emph{onboard queue} of a satellite as a waiting line where data units are held until they can be transmitted out or dropped due to limited storage capacity.

We assume the onboard queue is a first-in-first-out (FIFO) queue; in other words, the data in the queue is transmitted or dropped in the order of arrival.
For the satellite-ground station assignment, ground station access for low-priority satellites are prioritized based on proximity and assigned using the Hungarian algorithm~\cite{kuhn1955hungarian}, which is most commonly used in assigning communication channels between satellites and ground stations~\cite{vasisht2021l2d2,tao2023transmitting}; satellites closer to the ground station within a time slot receive superior access, ensuring more stable communication with less signal attenuation.
High-priority satellites, when contending for a ground station with low-priority satellites, are given precedence to utilize any idle antenna. In the absence of an available idle antenna, high-priority satellites are authorized to occupy antennas initially assigned to low-priority satellites. 
Consequently, displaced low-priority satellites are unable to use the occupied antennas until they become available again.
The transmissible time slot set of satellite $s_i$ is denoted as $\mathcal{X}_{s_i}$, which is the set of time slots during which the satellite $s_i$ has the opportunity to transmit data to one of the ground stations.
Note that, while we describe our attacks using a typical queue model (FIFO) and satellite-ground station assignment scheme, 
\purple{our attacks can also be applied to other types of queue models and assignment algorithms, as long as these are known to the attacker.}

\subsection{Threat Model}

\noindent Given a set of target data, denoted as $\Theta$, the data consists of a series of images, video fragments, or other types of data that can be captured and transmitted by a target low-priority satellite $s^* \in S_l$ at a certain time and location. We consider an active attacker seeking to either delay the downlink of $\Theta$ or drop $\Theta$ before it reaches the ground.

The attacker can only use legitimate services provided by the high-priority EO constellation, which allow their customers to collect and downlink data according to their interests. 
The attacker cannot access any other undisclosed information or functions. Specifically, the attacker knows all satellites' orbit dynamics, the location and configuration of shared ground stations, and the data capture pattern, image size, and average downlink rate of low-priority satellites from public sources; we justify these assumptions in \S\ref{sec:practical-consideration}.

The attacker deliberately schedules downlink tasks for the satellites in the high-priority constellation, causing them to occupy $s^*$'s originally allocated downlink resource at certain time slots, thereby disrupting the data transmission of $s^*$ and delaying the downlink of $\Theta$. With $s^*$ continuously generating new data and onboard data being queued by the attacker's designed strategy, 
\purple{$\Theta$ can be deliberately dropped.}
\purple{We consider the most challenging scenario for the attacker, where high-priority satellites lack data to preempt $s^*$'s downlink resource, as natural preemption would lower the attacker's cost.}

All time slots that the attacker decides to attack $s^*$ constitute the attack strategy $\mathcal{Y}_{s^*}$.
A time slot $t$ is ``attackable'' if there are a sufficient number of high-priority satellites available to occupy $s^*$'s downlink resource. 
All the attackable time slots of $s^*$ form the attackable time slot set $\mathcal{A}_{s^*}$.
Fig.~\ref{fig:attack_model} shows an example of the transmissible and attackable status for 4 low-priority satellites in one time slot.
    Satellite A cannot transfer data out since no ground station is available, making the current time slot neither transmissible nor attackable to it. Satellite B connects to one ground station, with only one high-priority satellite competing, making it transmissible but not attackable. 
    \purple{Satellites C and D can connect to a ground station, making the time slot transmissible, but competition from high-priority satellites makes it attackable.}

\blue{
    The cost of attacking a time slot $t$ is denoted as $\rho_{s^*}(t)$. Different cost functions can be applied depending on how services are priced in the real world and/or the amount of additional resources (such as fake user accounts) an attacker requires to launch an attack. In our evaluation, we consider costs that are proportional to the number of high-priority satellites involved, assuming that scheduling a service request for a particular satellite at a given time has a fixed price.}
The costs for all time slots are denoted as $\bm{\rho}_{s^*}$.

Overall, the attacker can conduct two types of attacks:
\vspace{-1mm}
\begin{itemize}

\item \noindent \textbf{Data Delay Attack.} The attacker aims to prevent the target data from being downlinked to the ground station until a specified later time, with minimum cost.

\item \noindent \textbf{Data Overflow Attack.} The attacker aims to target data being dropped due to the limited onboard queue capacity.
\end{itemize}


\subsection{Queue Evolution Model}
\noindent As mentioned in \S\ref{subsec:constellation_model}, we consider a FIFO queue for the onboard data.
Since data is collected on a per data unit basis, we use $|\Theta|$ to represent the number of data units in $\Theta$ and 
use $\tau \!\in\! \Theta$ to denote the data unit in $\Theta$ that is captured at time $T(\tau)$.
Specifically, $\tilde{\tau} \triangleq \arg \max_{\tau \in \Theta} \{T(\tau)\}$ denotes the last data unit in $\Theta$.
The onboard queue of $s^*$ at time slot $t$ under the attack strategy $\mathcal{Y}_{s^*}$ is denoted as $\mathcal{Q}_{s^*}(t,\mathcal{Y}_{s^*})$ with overall length $Q_{s^*}(t,\mathcal{Y}_{s^*})$, 
and the sub-queue including and preceding  $\tau$ is denoted as $\mathcal{Q}_{s^*}(\tau,t,\mathcal{Y}_{s^*})$ with length $Q_{s^*}(\tau,t,\mathcal{Y}_{s^*})$.
Given an attack start time $t_0 \!\in\! \mathbb{T}$, the state of the EO constellations 
with respect to $\Theta$ without any attack for  $s^*$ can be represented as $\{(\mathcal{X}_{s^*},\mathcal{A}_{s^*},\mathcal{Q}_{s^*}(t_0,\emptyset),$ $\mathcal{Q}_{s^*}(\tau,t_0,\emptyset),c_{s^*}), \tau \in \Theta\}$.

To model the two ways data is consumed in the onboard queue, we define the data amount downlinked to the ground station and the dropped data amount at time slot $t$ from $\mathcal{Q}_{s^*}(t,\mathcal{Y}_{s^*})$ as $O_{s^*}(t,\mathcal{Y}_{s^*})$ and $D_{s^*}(t,\mathcal{Y}_{s^*}) \!\triangleq\! \max \{0, Q_{s^*}(t-1,\mathcal{Y}_{s^*}) + I_{s^*}(t) - O_{s^*}(t,\mathcal{Y}_{s^*}) - c_{s^*}\}$, respectively.
We round up $D_{s^*}(t,\mathcal{Y}_{s^*})$ to $\max_{t\in \mathbb{T}} \{O(t, \mathcal{Y}_{s^*})\}$ if $0 \!<\! D_{s^*}(t,\mathcal{Y}_{s^*}) \!<\! \max_{t\in \mathbb{T}} \{O(t, \mathcal{Y}_{s^*})\}$ to ensure that the delay time caused by the attack is per time slot.
A \emph{data overflow} occurs at time slot $t$ when the onboard queue reaches its full capacity and has to drop some data, i.e., $ D_{s^*}(t,\mathcal{Y}_{s^*}) > 0$.

The queue length at time slot $t$ of $s^*$ is updated as $$ Q_{s^*}(t,\mathcal{Y}_{s^*}) \!=\! \min \{c_{s^*}, Q_{s^*}(t-1,\mathcal{Y}_{s^*}) + I_{s^*}(t) - O_{s^*}(t,\mathcal{Y}_{s^*})\}.$$

The sub-queue length at time slot $t$ of $s^*$ is 
$$Q_{s^*}(\tau, t, \mathcal{Y}_{s^*}) \!=\! \max \{0, Q_{s^*}(\tau, t_0, \emptyset) -\!\! \sum\nolimits_{t'=t_0}^{t} \Delta_{s^*}(t',\mathcal{Y}_{s^*})\},$$
where $\Delta_{s^*}(t,\mathcal{Y}_{s^*}) \!\triangleq\! O_{s^*}(t,\mathcal{Y}_{s^*}) \!+\! D_{s^*}(t,\mathcal{Y}_{s^*})$.

To describe how the attack strategy affects the final downlink time of $\tau$, we define the \emph{expected downlink time} of $\tau$ under attack strategy $\mathcal{Y}_{s^*}$ as
$$t_e(\tau, \mathcal{Y}_{s^*}) \triangleq \min_t \{Q_{s^*}(\tau,t,\mathcal{Y}_{s^*}) = 0 \}.$$
If $\tau$ is dropped due to the attack strategy $\mathcal{Y}_{s^*}$, then $t_e(\tau, \mathcal{Y}_{s^*}) \!=\! \infty$.
\purple{We define a metric called the \emph{attack strength} for a time slot $t'$ given an  attack strategy $\mathcal{Y}_{s^*}$ with respect to $\tau$ as $\Phi_{\tau}(\mathcal{Y}_{s^*}, t') \!\triangleq\! t_e(\tau, \mathcal{Y}_{s^*} \!\cup\! \{t'\}) \!-\! t_e(\tau, \mathcal{Y}_{s^*}).$}
This measures the additional delay when $t'$ is additionally attacked, compared to attacking only the time slots in $\mathcal{Y}_{s^*}$.
The overall attack strength of  $\mathcal{Y}_{s^*}$ for $\tau$ is $\Phi_{\tau}(\mathcal{Y}_{s^*}) \triangleq t_e(\tau, \mathcal{Y}_{s^*}) - t_e(\tau, \emptyset)$,
\purple{measuring the total delay for $\tau$ caused by $\mathcal{Y}_{s^*}$ compared to no attack.}




\begin{figure}[t]
    \centering\includegraphics[width=0.5\textwidth]{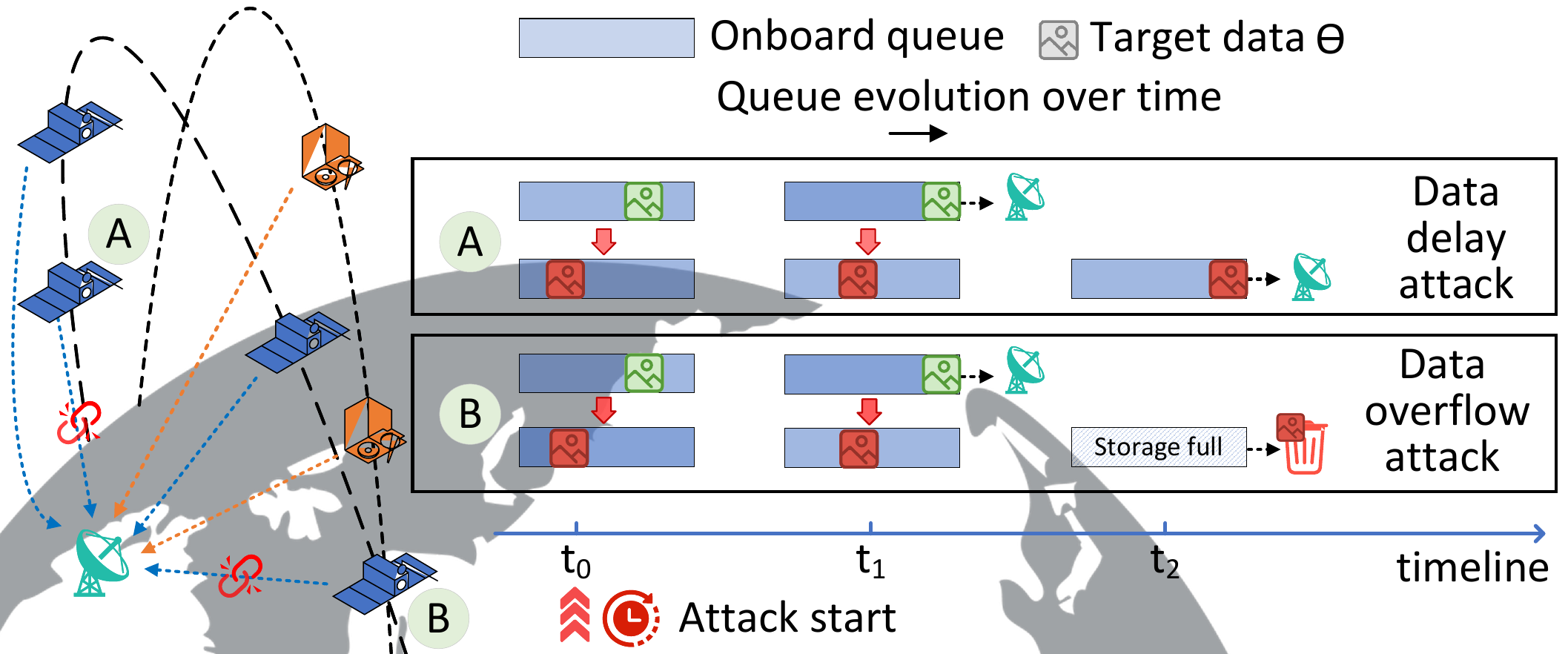}
    \vspace{1mm}
    \caption{Overview of data delay and data overflow attacks.}
    \label{fig:architecture}
\end{figure}

\section{Overview} 
\label{sec:overview}

\noindent In this section, we give an overview of data delay attack and data overflow attack.
The position of the target data in the onboard queue of satellite A under a data delay attack and in the onboard queue of satellite B under a data overflow attack is shown in Fig.~\ref{fig:architecture}.

The data delay attack is devised to prolong the delay of target data transmission from a low-priority satellite to the ground station. By targeting specific data and its corresponding satellite, the attacker schedules tasks to high-priority satellites, thereby occupying the time slots typically used by the target low-priority satellite for downlink. Consequently, the targeted data must wait longer for additional available time slots to downlink, resulting in significant delays in data retrieval and potential mission-critical impacts.
We design an algorithm to enable the attacker to find the minimum cost attack strategy in \S\ref{sec:delay-attack}. We show that when targeting unit data, our algorithm can always find a feasible attack strategy when one exists, and is optimal in terms of cost by proving its optimality when the queue is not empty upon initialization---an assumption typically met in practice due to limited communication resources.

The data overflow attack, exploiting the limited onboard capacity of satellites, poses a more severe threat than the data delay attack and leads to the dropping of the target data, rendering it irrecoverable.
As EO satellites continually collect new data, the objective of an attacker is to find an attack strategy that keeps the target data onboard until it is dropped when new data occupy its space.
This requires more delicate control over when and which time slots to attack, as attacking more slots may not result in a higher attack success probability. 
We devise an algorithm to search for the attack strategy, and for unit data as the target, our algorithm can return a feasible attack strategy whenever one exists.
See \S\ref{sec:buffer-overflow-attack} for more details.



\red{Proofs for the lemma and theorems of our theoretical analysis are provided in the Appendix.}



\section{Data Delay Attack}
\label{sec:delay-attack}
\begin{figure}[t]
    \centering
    \includegraphics[width=0.42\textwidth]{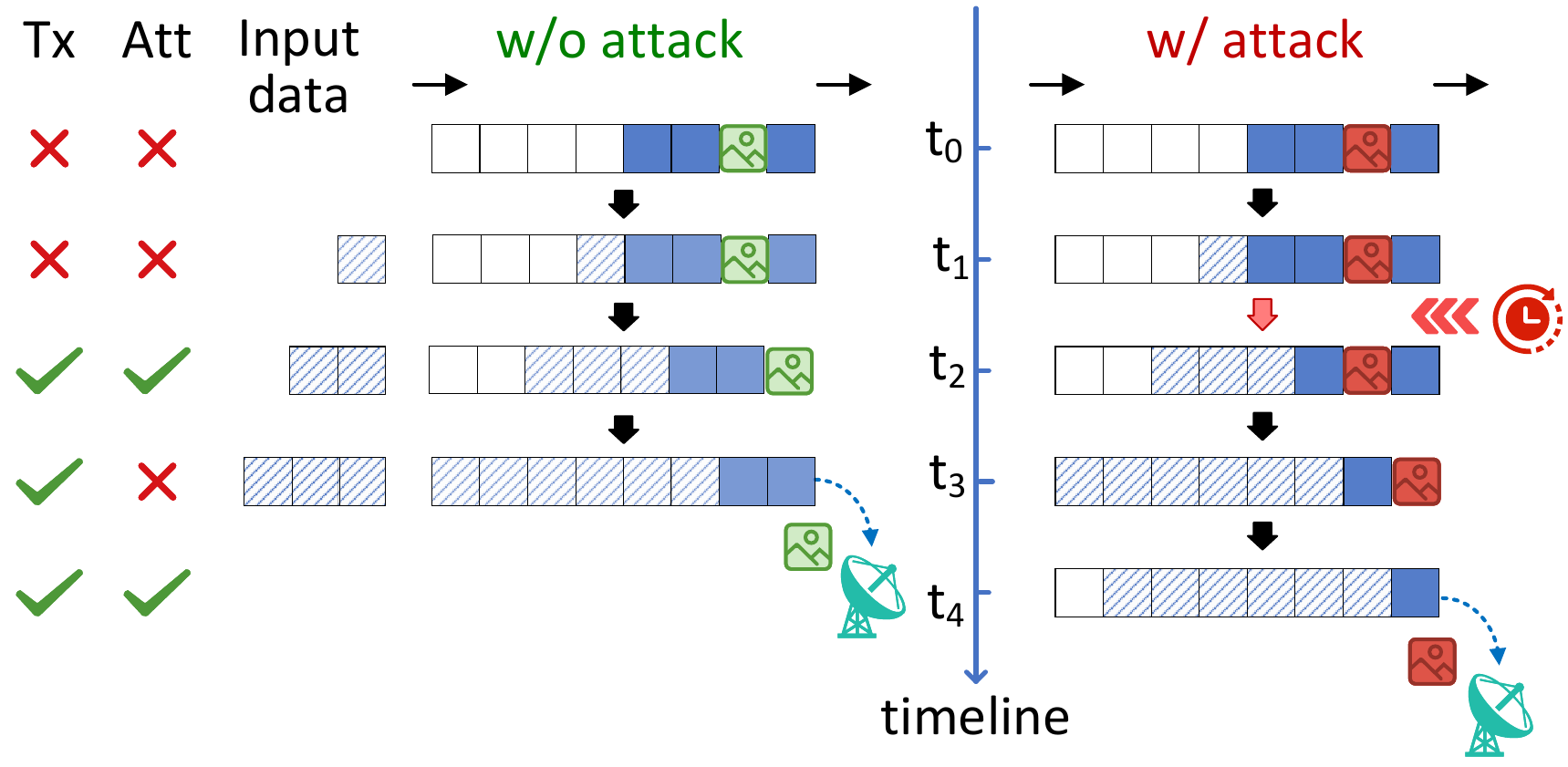}
    \vspace{2mm}
    \caption{
    The data delay attack.
    The timeline is from top to bottom, each block represents a data unit, and the queue evolves from left to right. 
    Without an attack, the target data would be downlinked at $t_3$. 
    With a delay attack at $t_2$, the target data is delayed one time slot and downlinked at $t_4$.
    }
    \label{fig:image-delay}
\end{figure}

\noindent An attacker aiming to execute a data delay attack seeks to prevent critical data on the target satellite from being downlinked to the ground station before a specific time. 
This data may involve real-time and highly urgent commands, such as those required for disaster response or key military operations. 
\purple{Delaying this data can cause confusion and disrupt crucial decisions.}
\purple{The attacker needs to choose suitable time slots and schedule high-priority satellite tasks to compete with the target satellite for ground station access.}
This competition results in the target satellite having no downlink resources available at the attacked time slots, forcing it to queue data onboard. 
The data queued before the target data takes longer to downlink, thereby delaying the target data’s downlink time.
An example of the data delay attack is shown in Fig.~\ref{fig:image-delay}.

\begin{figure}
    \centering\includegraphics[width=0.39\textwidth]{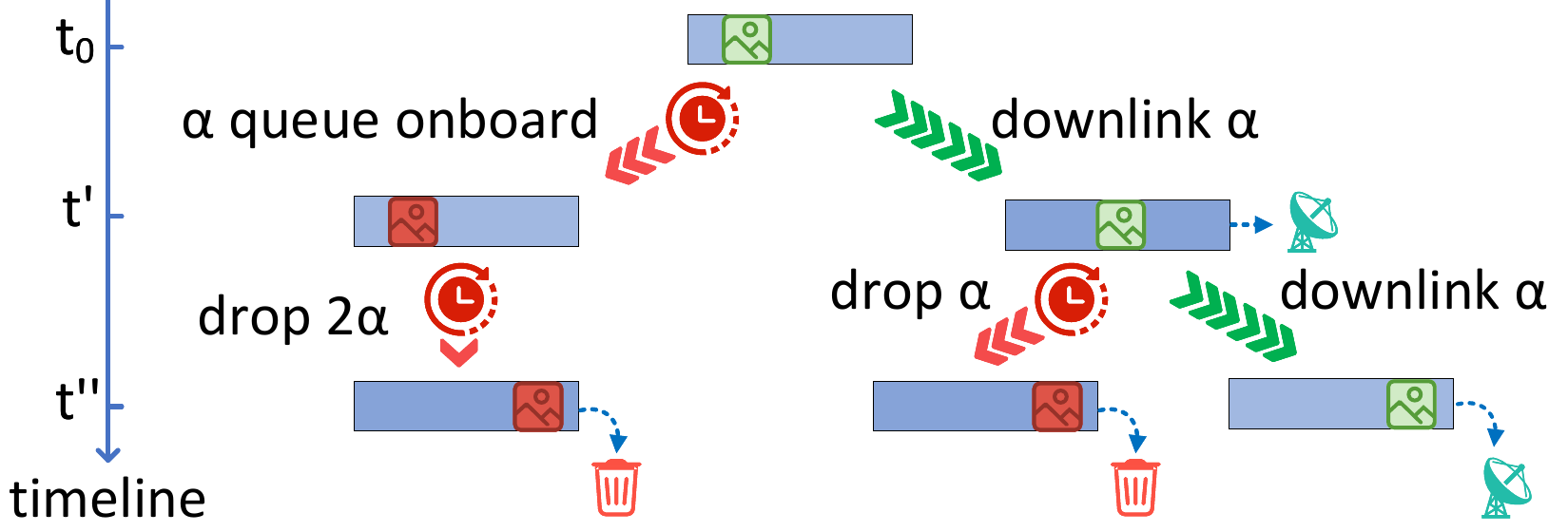}
    \vspace{2.5mm}
    \caption{
    Attacks before or at the point of queue full do not affect the target data's downlink time.
    Assume the attack at $t''$ causes a data overflow. 
    The data amount overflowed due to the attack equals the data amount meant to be downlinked during the attacked time slot; we denote this value as $\alpha$.
    }
    \label{fig:useless_example}
\end{figure}

An attacker's goal is to find an attack strategy that can delay the target data beyond a certain threshold, called the \emph{target downlink time} $t^*(\tilde{\tau}) > t_e(\tilde{\tau},\emptyset)$, with the minimum attack cost. We define the \emph{Data Delay problem} of finding the optimal data delay attack strategy as follows:

\begin{definition}
    \label{def:target-image-delay}
    Given the state of EO constellations $\{(\mathcal{X}_{s^*},\mathcal{A}_{s^*},\mathcal{Q}_{s^*}(t_0,\emptyset),$ $\mathcal{Q}_{s^*}(\tau,t_0,\emptyset),c_{s^*}), \tau \!\in\! \Theta\}$ at the attack start time $t_0$ for the target data $\Theta$ on satellite $s^*$, 
    the {Data Delay} problem is to identify an attack strategy $\mathcal{Y}_{s^*}$ that prevents $\Theta$ from being downlinked to the ground station until the target downlink time $t^*(\tilde{\tau})$, with minimum cost.
This can be formulated as follows:
\begin{align}
    \label{eq:delay-attack}
    &\min_{\mathcal{Y}_{s^*}} \quad \sum_{t \in \mathcal{Y}_{s^*}} \rho_{s^*}(t) \quad \text{s.t.} 
     \quad t_e(\tilde{\tau}, \mathcal{Y}_{s^*}) > t^*(\tilde{\tau}). \notag 
\end{align}
\end{definition}

Intuitively, if the satellite's internal storage is infinite, then attacking a transmissible time slot before the target data downlink will directly result in the target data's downlink time being delayed by one transmissible time slot.
However, what complicates the attack strategy is that if an attack results in the onboard queue being full at this time slot or some time later than the attacked slot but before the target downlink time, then further attacking any slot before or during the onboard queue being full will have no impact on further delaying the downlink time of the target data.
As shown in Fig.~\ref{fig:useless_example}, regardless of whether an attack occurs at $t'$ or not, by the end of $t''$, the amount of data ahead of the target data is the same as it would be if no attack had occurred up to $t''$.
To model the queue full condition,
we define $t_{lb}(\tau, \mathcal{Y}_{s^*}) \!\triangleq\! \max\{ t_0, \max_t \{Q_{s^*}(t,\mathcal{Y}_{s^*}) \!=\! c_{s^*} \text{ and } t_0 \!\le\! t \!<\! t_e(\tau, \mathcal{Y}_{s^*})\}\}$
as the last time slot when the queue is full before the data unit $\tau$ is transmitted under the attack strategy $\mathcal{Y}_{s^*}$.
If no data overflow occurs before ${\tau}$ is transmitted out, then $t_{lb}(\tau, \mathcal{Y}_{s^*}) = t_0$.

The following lemma shows that attacks on time slots at or before the queue is full do not delay the data unit $\tau \in \Theta$.

\begin{lemma}
    \label{lemma:buffer-overflow-useless}
    Given a data unit $\tau$, 
    an attack strategy $\mathcal{Y}_{s^*}$ and a set $\mathcal{A}_{s^*}(t_{lb}(\tau, \mathcal{Y}_{s^*})) \!\triangleq\! \{ t| t \in \mathcal{A}_{s^*} \text{ and } t_0 \!<\! t \!\le\! t_{lb}(\tau, \mathcal{Y}_{s^*})\}$,
    $t_e(\tau, \mathcal{Y}_{s^*} \!\cup\! \mathcal{A}') \!=\! t_e(\tau, \mathcal{Y}_{s^*})$ 
    for all $\mathcal{A}' \subseteq \mathcal{A}_{s^*}(t_{lb}(\tau, \mathcal{Y}_{s^*}))$.
\end{lemma}

\begin{algorithm} [t]
    \caption{Data Delay Attack} \label{al:target_image_delay}
    \LinesNumbered
    \KwIn{Target data $\Theta$, target downlink time $t^*(\tilde{\tau})$, EO constellations state $\{(\mathcal{X}_{s^*},\mathcal{A}_{s^*},\mathcal{Q}_{s^*}(t_0,\emptyset),$ $\mathcal{Q}_{s^*}(\tau,t_0,\emptyset),c_{s^*}), \tau \in \Theta\}$,
    attack cost set $\bm{\rho}_{s^*}$
    }
    \KwOut{Attack strategy $\mathcal{Y}_{s^*}$}

    $\mathcal{Y}_{s^*} \leftarrow \emptyset$\; \label{line:init_Y}

    \For{ $\tau \in \Theta$ }{

    $t_e(\tau,\mathcal{Y}_{s^*}) \leftarrow \min_t \{t|Q_{s^*}(\tau,t, \mathcal{Y}_{s^*}) = 0 \}$\;
    $t_{lb}(\tau, \mathcal{Y}_{s^*}) \leftarrow \max\{ t_0, \max_t \{t|Q_{s^*}(t,\mathcal{Y}_{s^*}) = c_{s^*} \text{ and } t_0 \le t < t_e(\tau)\}\}$\;\label{line:init_lb}

    \lIf{data unit $\tau$ is dropped}{\Return $\mathcal{Y}_{s^*}$\label{line:drop1}}
    
    {
    
        {
            
            \While{$t_e(\tau,\mathcal{Y}_{s^*}) - t_e(\tau,\emptyset) \le t^*(\tilde{\tau}) - t_e(\tilde{\tau},\emptyset)$ \label{line:repeat}}
            {
            
                $\hat{T}_\tau \leftarrow \{t | t_{lb}(\tau, \mathcal{Y}_{s^*}) < t \le t_e(\tau, \mathcal{Y}_{s^*}) \text{ and } t \in \mathcal{A}_{s^*} \text{ and } t \notin \mathcal{Y}_{s^*}\}$\; \label{line:sort_T}

                \lIf{$\hat{T}_\tau = \emptyset$}{\Return Attack Fail\label{line:attack_fail}}
                $\hat{t} \leftarrow \arg \min_{t \in \hat{T}_\tau} \{{\rho_{s^*}(t)}\}$\;\label{line:find_min_cost_time_slot}

                $\mathcal{Y}_{s^*} \leftarrow \mathcal{Y}_{s^*} \cup \{\hat{t}\}$\;\label{line:add_t'}
                $t_e(\tau,\mathcal{Y}_{s^*}) \leftarrow \min_t \{t|Q_{s^*}(\tau,t, \mathcal{Y}_{s^*}) = 0 \}$\;

                \If{$t_e(\tau,\mathcal{Y}_{s^*}) = \infty$ ($\tau$ is dropped)}
                {
                    \Return Attack strategy $\mathcal{Y}_{s^*}$\;\label{line:drop}
                }

            $t_{lb}(\tau, \mathcal{Y}_{s^*}) \leftarrow \max\{ t_0, \max_t \{t|Q_{s^*}(t,\mathcal{Y}_{s^*}) = c_{s^*} \text{ and } t_0 \le t < t_e(\tau,\mathcal{Y}_{s^*})\}\}$\;\label{line:update_lb}

            }
        }
        
    }
    }
    
    \Return Attack strategy $\mathcal{Y}_{s^*}$\;
\end{algorithm}

\purple{Based on Lemma~\ref{lemma:buffer-overflow-useless}, we devise Algorithm~\ref{al:target_image_delay} to find a feasible data delay attack strategy and prove its optimality for unit target data.}
We first initialize the attack strategy $\mathcal{Y}_{s^*}$. Then for each data unit $\tau \in \Theta$,  we get the expected downlink time and the last queue full time without any attack (lines~\ref{line:init_Y}-\ref{line:init_lb}).
If $\tau$ is dropped without any attack, then we do not need to conduct the delay attack (line~\ref{line:drop1}).
If the evacuation time $t_e(\tau,\mathcal{Y}_{s^*}) - t_e(\tau,\emptyset)$ is smaller than the target duration $t^*(\tilde{\tau}) - t_e(\tau, \mathcal{Y}_{s^*})$, which means $\tau$ needs more attack time slots to delay longer, we then begin to find the minimum cost attack strategy for $\tau$ (line~\ref{line:repeat}).
We first construct a set $\hat{T}_{\tau}$ that contains all the attackable time slots that potentially have attack strength for $\tau$ (line~\ref{line:sort_T}).
Then we find the minimum cost attack time slot in $\hat{T}_{\tau}$ as $\hat{t}$ (line~\ref{line:find_min_cost_time_slot}). 
{If multiple attackable slots have the same cost, the earliest one is chosen.}
If the set $\hat{T}_{\tau}$ is empty, indicating the absence of any time slot that can be attacked, then the attack fails and $\tau$ can be transmitted before $t^*(\tilde{\tau})$ no matter how the attacker attacks (line~\ref{line:attack_fail}).
If $\hat{t}$ is found, we add $\hat{t}$ to the attack strategy $\mathcal{Y}_{s^*}$ and update the expected downlink time $t_e(\tau,\mathcal{Y}_{s^*})$ and the last queue full time $t_{lb}(\tau, \mathcal{Y}_{s^*})$ (lines~\ref{line:add_t'}-\ref{line:update_lb}).
These two times need to be updated because additional attacks on $\hat{t}$ might influence whether further exploration of new attack time slots is needed and exclude time slots that definitely do not have attack strength.
\purple{We repeat the process until the delay target is met (line~\ref{line:repeat}), or if the target data is dropped (line~\ref{line:drop}).}

    \purple{EO satellites operate on predefined orbits, covering large regions of the Earth at fixed times.}
    If the attacker’s goal is to delay or drop data of a specific location at a specific time, due to imaging frequency limitations, only a few or possibly just one data unit may be a uniquely valuable resource.
    \purple{In the following, we analyze the completeness and optimality of Algorithm~\ref{al:target_image_delay} in finding the minimum cost attack strategy for single-unit target data, which suffices for many attack tasks.}

\begin{theorem}
\label{theorem:target-image-delay-attack-optimal}
    \red{When $|\Theta|=1$ and $Q_{s^*}(t, \emptyset) > 0$ for all $t \in \mathbb{T}$, if there exists a feasible solution that can delay the target data until the target downlink time,  then Algorithm~\ref{al:target_image_delay} will always terminate with a feasible attack strategy which has the minimum cost for the {Data Delay} problem.}
\end{theorem}



\begin{figure}[t]
    \centering
    \includegraphics[width=0.42\textwidth]{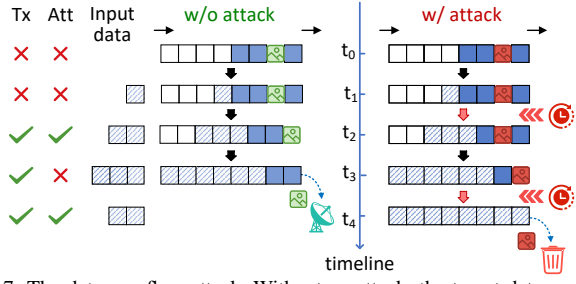}
    \vspace{2mm}
    \caption{The data overflow attack. Without an attack, the target data would be downlinked to the ground station at $t_3$. After attacks at $t_2$ and $t_4$, the target data is irretrievably dropped at $t_4$.}
    \label{fig:buffer_overflow_attack}
\end{figure}
\section{Data Overflow Attack}
\label{sec:buffer-overflow-attack}

\noindent For the data overflow attack, given target data $\Theta$, the goal of the attacker is to induce data overflow on the target satellite $s^*$ so that the $\Theta$ is dropped. 
The attacker disrupts the downlink process of $s^*$ by blocking specific attackable time slots, consequently delaying the expected downlink time for all data that is scheduled to be downlinked in subsequent slots. 
With the satellite continuously capturing new data and the attacker-caused delays resulting in additional data stored onboard, the limited onboard storage will eventually fill up, causing a data overflow.
As the input data of the onboard queue is not controlled by the attacker, the key point of the data overflow attack is that the attacker needs to ensure that the target data remains stored onboard until a data overflow occurs, and that when the data overflow happens, $\Theta$ is at the head of the queue, thereby ensuring that $\Theta$ can be dropped.
Fig.~\ref{fig:buffer_overflow_attack} gives an example of data overflow attack.

An attacker's goal is to find a feasible attack strategy to drop the onboard target data, preventing it from being downlinked to the ground station.
We define the \emph{Data Overflow problem} of finding the feasible data overflow attack strategy as follows:

\begin{definition}
    \label{def:buffer-overflow}
   Given the state of EO constellations $\{(\mathcal{X}_{s^*},\mathcal{A}_{s^*},\mathcal{Q}_{s^*}(t_0,\emptyset),$ $\mathcal{Q}_{s^*}(\tau,t_0,\emptyset),c_{s^*}), \tau \!\in\! \Theta\}$ at the attack start time $t_0$ for the target data $\Theta$ on satellite $s^*$, 
    the {Data Overflow} problem is to find an attack strategy $\mathcal{Y}_{s^*}$ that satisfies the following conditions: $Q_{s^*}(\tilde{\tau},t,\mathcal{Y}_{s^*}) = 0 \text{ and } Q_{s^*}(t,\mathcal{Y}_{s^*}) = c_{s^*} \text{ for some } t \ge t_0$.
\end{definition}

\begin{algorithm}[t]
    \caption{Data Overflow Attack} \label{alg:buffer_overflow}
    \LinesNumbered
    \KwIn{Target data $\Theta$, EO constellations state $\{(\mathcal{X}_{s^*}, $ $\mathcal{A}_{s^*}, \mathcal{Q}_{s^*}(t_0,\emptyset),$ $\mathcal{Q}_{s^*}(\tau,t_0,\emptyset),c_{s^*}), \tau \in \Theta\}$
    }
    \KwOut{Attack strategy $\mathcal{Y}_{s^*}$}

    $\mathcal{Y}_{s^*} \leftarrow \emptyset$\; \label{line:init_attack_set_bf}
    \For{ $\tau \in \Theta$ }{
    $t_e(\tau,\mathcal{Y}_{s^*}) \leftarrow \min_t \{t|Q_{s^*}(\tau,t, \mathcal{Y}_{s^*}) = 0 \}$\; \label{line:init_target_image_downlink_time_bf}
    $t_{lb}(\tau, \emptyset) \leftarrow \max\{ t_0, \max_t \{Q_{s^*}(t_0) = c_{s^*} \text{ and } t_0 \le t < t_e(\tau,\mathcal{Y}_{s^*})\}\}$\;\label{line:init_last_buffer_overflow_time_bf}
    Sort set $\{t| t \ge t_e(\tau,\mathcal{Y}_{s^*}) \text{ and } t \in \mathcal{A}_{s^*}\}$ in ascending order as $\textsf{T}_{n}$\;\label{line:sort_next_attack_time_slot_bf}

    Sort set $\{t| t_0 \le t < t_e(\tau,\mathcal{Y}_{s^*}) \text{ and } t \in \mathcal{A}_{s^*}\}$ in descending order as $\textsf{T}_{p}$\;\label{line:sort_previous_attack_time_slot_bf}
    $t_n \leftarrow \textsf{T}_n.pop()$, $t_p \leftarrow \textsf{T}_{p}.pop()$\;\label{line:pop_out_tn_tp_bf}

    \While {$\tau$ is not be dropped}{
        \If {$t_n \le t_e(\tau, \mathcal{Y}_{s^*})$ \label{line:attack_tn_bf}}{
            $\mathcal{Y}_{s^*} \leftarrow \mathcal{Y}_{s^*} \cup \{t_n\}$\;
            
            $t_e(\tau,\mathcal{Y}_{s^*}) \leftarrow \min_t \{t|Q_{s^*}(\tau,t, \mathcal{Y}_{s^*}) = 0 \}$\;
            
            $t_n \leftarrow \textsf{T}_{n}.pop()$\;\label{line:update_tn_bf}

        }

            \ElseIf{\purple{$\textsf{T}_p \neq \emptyset$} and $t_p > t_{lb}(\tau,\mathcal{Y}_{s^*})$\label{line:attack_tp_bf}}{
            $\mathcal{Y}_{s^*} \leftarrow \mathcal{Y}_{s^*} \cup \{t_p\}$\;
            $t_e(\tau,\mathcal{Y}_{s^*}) \leftarrow \min_t \{t|Q_{s^*}(\tau,t, \mathcal{Y}_{s^*}) = 0 \}$\;

            $t_{lb}(\tau, \mathcal{Y}_{s^*}) \leftarrow \max\{ t_0, \max_t \{t|Q_{s^*}(t,\mathcal{Y}_{s^*})$ $ = c_{s^*} \text{ and } t_0 \le t < t_e(\tau)\}\}$\;
            $t_p \leftarrow \textsf{T}_{p}.pop()$\; \label{line:update_tp_bf}

        }
        \lElse {\Return Attack Fail\label{line:attack_fail_bf}}
    }
    }
       
    \Return Attack strategy $\mathcal{Y}_{s^*}$.
\end{algorithm}

\purple{We design Algorithm~\ref{alg:buffer_overflow} to solve the {Data Overflow} problem.}
The attack begins by initializing the attack strategy $\mathcal{Y}_{s^*}$ as an empty set (line~\ref{line:init_attack_set_bf}).
\purple{For each data unit $\tau$, we calculate its expected downlink time $t_e(\tau,\mathcal{Y}_{s^*})$ and the initial queue full time $t_{lb}(\tau, \emptyset)$ without any attack (lines~\ref{line:init_target_image_downlink_time_bf}$-$\ref{line:init_last_buffer_overflow_time_bf}).}
In addition, we generate two sets of time slots: $\textsf{T}_{n}$ is the set of attackable time slots at and after the initial image downlink time that are sorted in ascending order, and $\textsf{T}_{p}$ is the set of attackable time slots before the initial image downlink time that are sorted in descending order (lines~\ref{line:sort_next_attack_time_slot_bf}$-$\ref{line:sort_previous_attack_time_slot_bf}).
$t_n$ and $t_p$ are the next attack time slot in $\textsf{T}_{n}$ and $\textsf{T}_{p}$, respectively (line~\ref{line:pop_out_tn_tp_bf}).
If $t_n \!\le\! t_e(\tau,\mathcal{Y}_{s^*})$, which means we can attack the time slot $t_n$ and keep the target data onboard and at the top of the queue, then we update the expected downlink time $t_e(\tau,\mathcal{Y}_{s^*})$ based on the new attack strategy $\mathcal{Y}_{s^*}$ that includes the time slot $t_n$ and update $t_n$ to the next element in $\textsf{T}_{n}$ (lines~\ref{line:attack_tn_bf}$-$\ref{line:update_tn_bf}).
Else if $t_n > t_e(\tau,\mathcal{Y}_{s^*})$, which means the next attackable time slot in $\textsf{T}_{n}$ cannot contribute to delaying the downlink time of the target data, we need to attack the time slot $t_p$ in $\textsf{T}_{p}$ to delay the downlink time of $\tau$ to make sure $\tau$ remains onboard (lines~\ref{line:attack_tp_bf}$-$\ref{line:update_tp_bf}).
\purple{If $\textsf{T}_p$ is empty or new data overflow happens due to the attack on $t_p$, then the attack fails (line~\ref{line:attack_fail_bf}).}
This means there is no available attackable time slot to delay $\tau$ further; $\tau$ is downlinked before it can be dropped.
\purple{This process repeats for each target unit in $\Theta$ until all are dropped.}
\purple{Theorem~\ref{theorem:buffer-overflow-attack-algorithm} shows the completeness of Algorithm~\ref{alg:buffer_overflow} for targeting a single data unit.}

\begin{theorem}
    \label{theorem:buffer-overflow-attack-algorithm}

        When $|\Theta|\!=\!1$, Algorithm~\ref{alg:buffer_overflow} terminates with a feasible attack strategy for the {Data Overflow} problem whenever
        a feasible solution exists.
\end{theorem}



\vspace{-1.5mm}
\section{Practical Considerations}
\label{sec:practical-consideration}
\vspace{-1mm}
\noindent
\dgreen{In this section, we analyze the feasibility of the attacks and discuss noise tolerance in real-world scenarios.}
\begin{figure}
    \centering
    \includegraphics[width=0.5\textwidth]{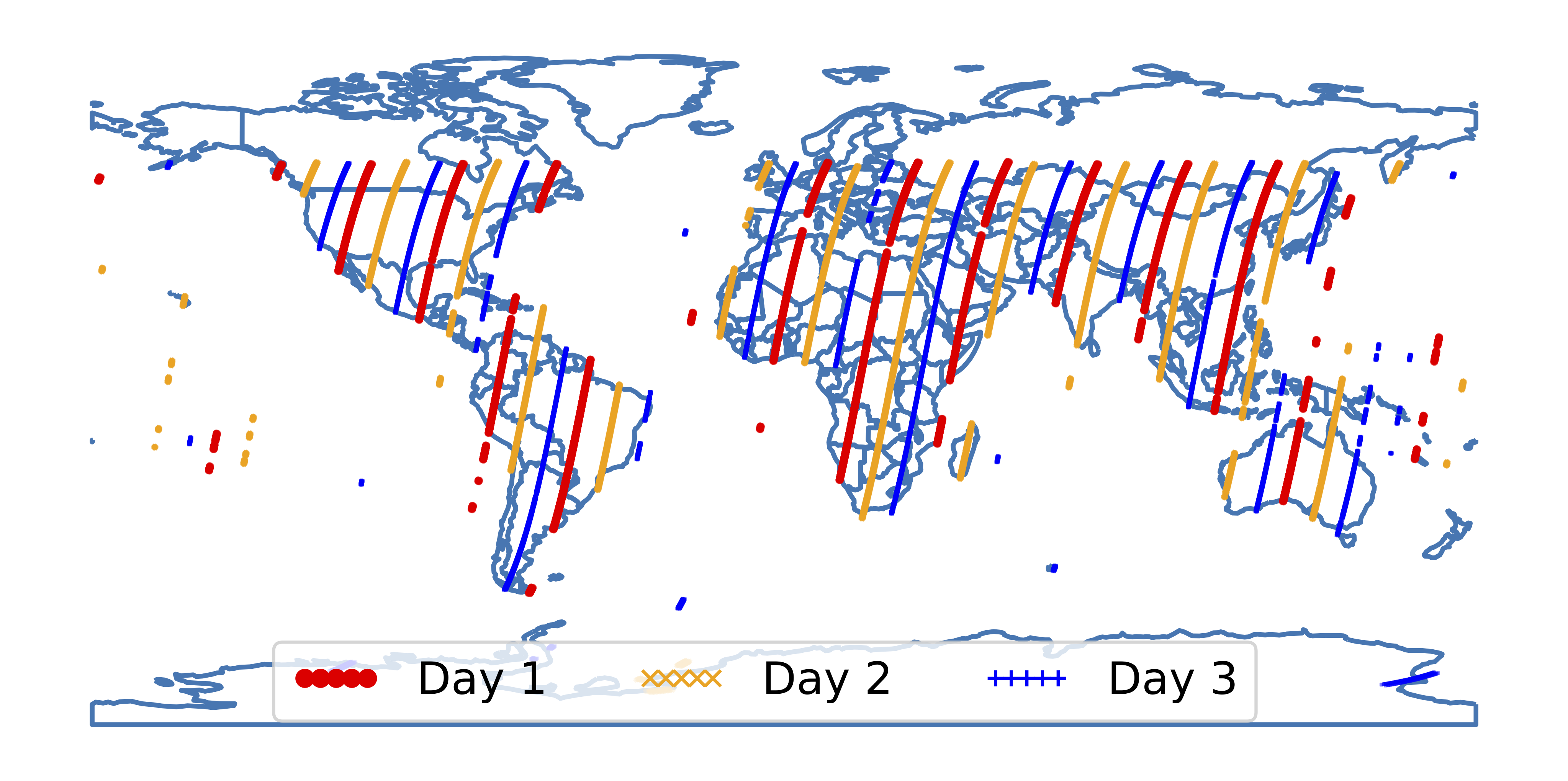}
    \caption{Dove 2473 satellite geographical 
 trace from January 5 to 7, 2024.}
    \label{fig:satellite-trace}
\end{figure}

\dgreen{Our attacks' feasibility depends on three main conditions: (1) contention for limited communication resources between high- and low-priority satellites,}
\gray{(2) the attacker's ability to schedule transmission of high-priority tasks for specific satellites at specific times,}
\dgreen{and (3) the attacker's knowledge of orbit information, scheduling policies, and the target satellite's queue status. We next explain why these conditions are very likely met by current and future constellations.}

\dgreen{\textbf{(1) Resource contention.}
Given the high cost and environmental requirements for building ground stations, sharing of ground stations among multiple constellations are becoming increasingly common for minimizing cost and maximizing performance~\cite{ground_station_services, aws_ground_station}.
For instance, the Ground Station Operations team at Planet Labs Inc.\ has begun integrating software and hardware architectures across different constellations to support multiple missions through their ground station networks, aiming to reduce on-orbit reaction latency and increase the utilization of individual ground stations~\cite{colton2020merging}.%
}

\revision{\textbf{(2) Attacker's ability to schedule transmission of high-priority tasks.}
Critical applications such as rapid disaster response or real-time emergency management require the delivering of user-scheduled high-priority, time-sensitive sensing data to the ground, which is being supported by more and more EO constellations.
Such services can then be utilized by an attacker to preempt low-priority transmission during shared communication windows.
For instance, 
Planet Labs allows users to schedule \emph{Assured Tasking Orders} at specific times when satellites pass over designated areas~\cite{assured_tasking,task_order}, and have the data delivered with minimum delay via its \emph{Fast Track Delivery} service~\cite{Fast_Track_Delivery}.
The data of such orders will be transmitted as soon as possible in the next communication windows after being generated, which opens up a possible avenue for adversarial scheduling of such orders to launch an attack.
The attacker can further profile the delay in launching an attack, by submitting multiple high-priority tasks and observing their downlink/delivery times prior to launching the actual attack.
}

\dgreen{\textbf{(3) Attacker's knowledge of satellite orbits, scheduling, and queues.}
Firstly, the orbit information for both low- and high-priority satellites is publicly accessible~\cite{real_TLE}, and since they must adhere to the orbit parameters claimed before their launch in FCC regulations, their future orbits are also predictable.
Secondly, an attacker could obtain the communication scheduling policy via various methods. 
For instance, there has been extensive research on scheduling policy inference from measurement data, using deep learning~\cite{chen2021deep} or probabilistic inference~\cite{gehr2018bayonet}.
An attacker could also compromise a satellite (possibly even an outdated or decommissioned one), a ground station, or even a cloud-hosted copy of the satellite communication codebase, to uncover the scheduling policy.
An insider attack from within the operator's organization is also possible.
Thirdly,}
an attacker can use various information sources to assist in determining the initial queue's data volume.
\purple{Fig.~\ref{fig:satellite-trace} shows the data collection pattern of a single satellite over three days.}
The satellite's data collection is predictable and periodically repeats, focusing solely on gathering data above specific areas of the Earth's surface.
For a specific satellite, the dynamics of data generation rate and captured images can be reliably predicted.
So if an attacker can determine the size of the queue at a certain moment and utilize the queue's data generation and downlink patterns, it can relatively accurately determine the initial queue size at the start of the attack.

\purple{To determine the queue size at a specific time, an attacker with access to the satellite and ground station channel can infer that a satellite's queue is empty if it does not downlink data during a transmissible window with no competition.}
\purple{Without channel access, the attacker can infer this from publicly available data via the open-source global network of satellite ground stations~\cite{crowdsourcing-ground-station}.}
Other starting points for the attacker include evolving the queue from the target satellite's initial launch.
\dgreen{Other methods could also be used to infer queue status at a specific time, for instance, using input/output communication patterns with fuzzy inference systems~\cite{tan2024reinforcement} or other queue length estimation algorithms~\cite{chan2013queue}.}

\dgreen{\textbf{Dealing with estimation noise.}}
\dgreen{An attacker may mis-estimate data size and data rate in the real-world, which may lead to failure of attacks if they are calculated based on inaccurate information. Heuristic strategies can be employed to deal with noise in such estimation.}
For instance,
the attacker can increase the attack's robustness against noise by attacking $M$ extra data units both before and after its primary target of $N$ units. 
If it can find an attack strategy that is feasible for all $N\!+\!2M$ data units as the target, then even if noise in queue dynamics or data rate estimation causes the attack to be unsuccessful on some data at the beginning or the end, there is still a high chance that the true target of $N$ units is successfully attacked.
\purple{We show that this strategy can significantly increase both attacks' robustness to estimation noise in \S\ref{sec:eval}.}

There are other approaches to mitigate the impact of noise.
For example, in a data delay attack, the attacker can generate a strategy to delay the target data past the target downlink time by a few time slots.
Alternatively, redundantly attacking multiple time slots near the target data's downlink time can also help combat noise.
Since attack strategies can be generated and tested through simulation, the attacker can tailor multiple strategies for specific target data and select the most robust one against noise for a real-world attack.



\section{Performance Evaluation}
\label{sec:eval}
\subsection{Experiment Settings}

\purple{\noindent We conducted a trace-driven simulation to evaluate the proposed attacks.}
We obtained the real-world metadata of the Planet Labs Dove satellite image data for January 2024 from the Planet API~\cite{Planet_API}. 
We selected the top 10 Dove satellites with the highest data generation on January 1st from the 118 satellites collecting data in January 2024 as the target satellites. From these satellites, we randomly sampled 1000 images (100 images from each satellite) among the total of 321,744 images generated by the entire Dove constellation.

Currently, the SkySat constellation consists of 21 satellites. 
As more satellites are planned for launch in the future, we extended the high-priority constellation to include 50 satellites to explore potential future attack performance.
The additional 29 high-priority satellites were selected from other companies' EO constellations~\cite{EO_pool}, prioritizing those with the most overlap in ground station sharing with Dove satellites. We also assessed the attack performance by varying the number of high-priority satellites, with the default setting as 50.

We utilized real-world orbit Two Line Elements (TLEs) information~\cite{real_TLE} of all the satellites to simulate the satellite constellation's orbital dynamics.
Twelve ground stations are distributed around the world, each equipped with 4 antennas, with a visibility threshold set at 5 degrees~\cite{tao2023transmitting,48-ground-stations}. 
\orange{The average downlink rate from satellite to ground station was set to 160Mbit/s, aligned with the real-world settings~\cite{Planet_downlink_rate}.}

\orange{The onboard storage of a satellite is mainly limited by volume and weight, and is non-upgradable after launched to the space.}
\orange{We set the onboard storage to be 2000GB, which is aligned with the existing Dove satellite~\cite{tao2023transmitting}.}
We used image as the type of data captured by the satellite; the data could be any other type such as hyperspectral, radar, etc.
The default settings had an image size of 200MB, $|\Theta|\!=\!4$ for 4 images as target data, 
an initial queue size of 500 images, and a data rate of 160Mbit/s. 
We varied parameters such as image size (ranging from 200MB to 500MB), data rate (ranging from 80Mbit/s to 320Mbit/s), high-priority satellites number (ranging from 21 to 50), and cost budget (ranging from 500 to 4000)
 to evaluate the impact of these factors on the attack performance.
Gaussian noise was added to simulate real-world conditions where the attacker may not have precise knowledge of the image size that may vary due to the nature of the data collected by the satellite or the different types of tasks, or the data rate that fluctuates due to the weather or other factors. 
The default setting had a standard deviation that was $0.1$ times the true value, and we also varied this ratio from $0$ to $0.4$.
To simulate the scenario where the attacker lacks precise information regarding the initial queue size, we incorporated random noise of $\pm$10 images for the initial queue.
For the data delay attack, we explored delay times ranging from 1 hour to 24 hours, with 24 hours as the default value.
Each experiment was conducted with 10 different seeds to mitigate randomness.

\begin{figure}[t]
    \centering
    \begin{subfigure}[b]{0.23\textwidth}
        \centering
        \includegraphics[width=\textwidth]{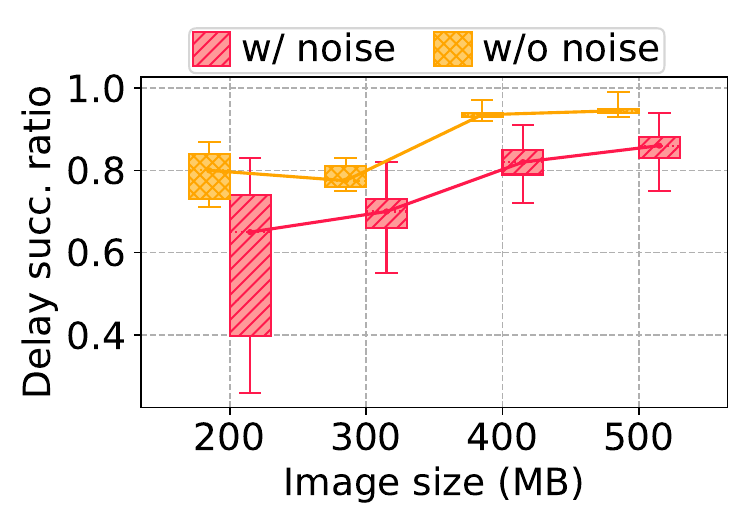}
        \label{fig:delay_success_ratio_vary_image_size}
    \end{subfigure}
    \begin{subfigure}[b]{0.23\textwidth}
        \centering
        \includegraphics[width=\textwidth]{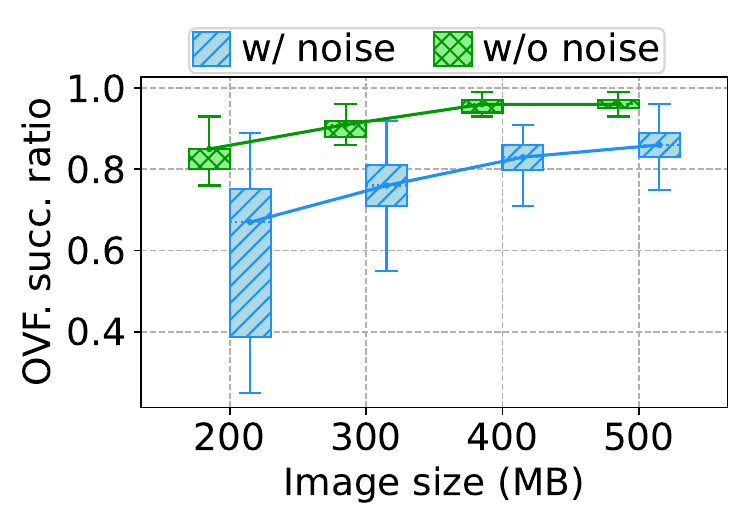}
        \label{fig:BF_success_ratio_vary_image_size}
    \end{subfigure}
    \vspace{-2.5mm}
    \caption{Attack success ratios vs. image size.}
    \label{fig:success_ratio_vs_image_size}
    \vspace{-0.5mm}
\end{figure}

\begin{figure}[t]
    \centering
    \begin{subfigure}[b]{0.23\textwidth}
        \centering
        \includegraphics[width=\textwidth]{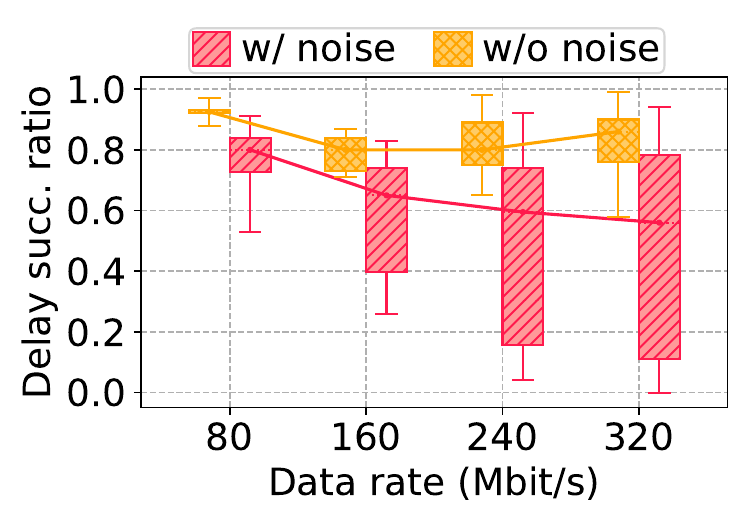}
        \label{fig:delay_success_ratio_vary_data_rate}
    \end{subfigure}
    \begin{subfigure}[b]{0.23\textwidth}
        \centering
        \includegraphics[width=\textwidth]{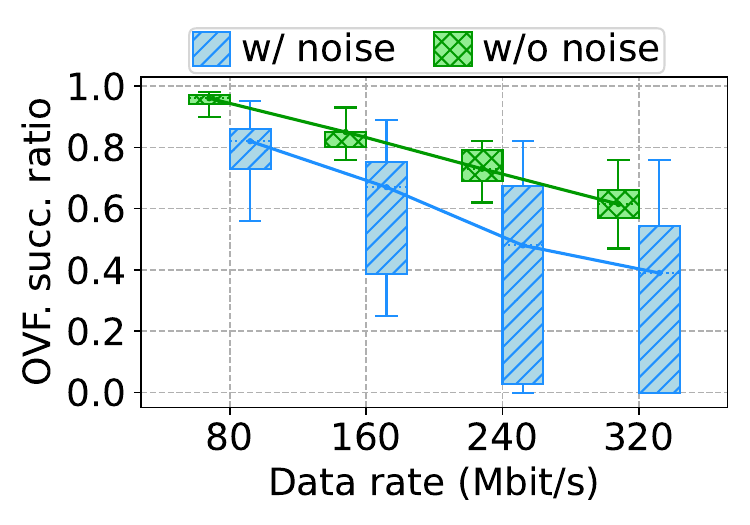}
        \label{fig:BF_success_ratio_vary_data_rate}
    \end{subfigure}
    \vspace{-2.5mm}
    \caption{Attack success ratios vs. data rate.}
    \label{fig:success_ratio_vs_data_rate}
    \vspace{-0.5mm}
\end{figure}

\begin{figure}[t]
    \centering
    \begin{subfigure}[b]{0.23\textwidth}
        \centering
        \includegraphics[width=\textwidth]{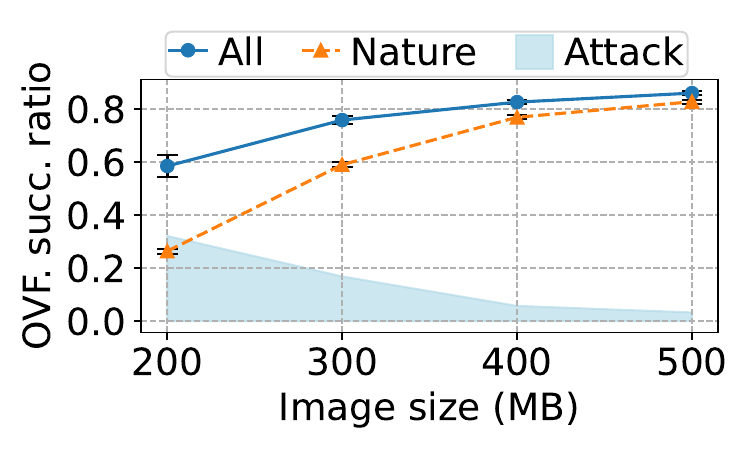}
        \label{fig:BF_need_attack_vary_image_size}
    \end{subfigure}
    \begin{subfigure}[b]{0.23\textwidth}
        \centering
        \includegraphics[width=\textwidth]{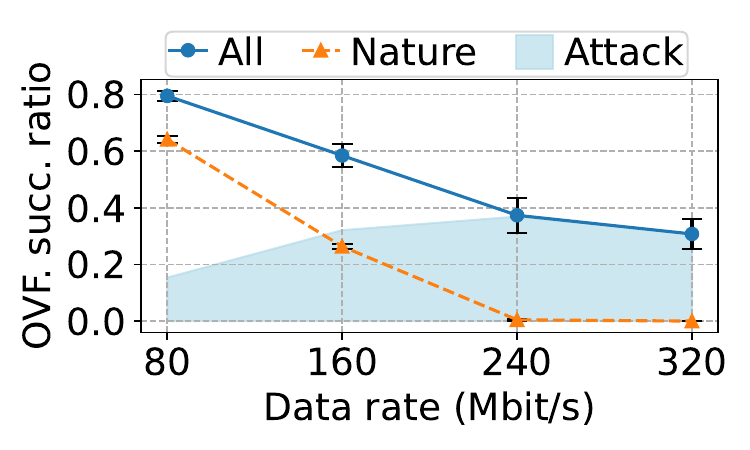}
        \label{fig:BF_need_attack_vary_data_rate}
    \end{subfigure}
    \vspace{-2.5mm}
    \caption{Data overflow attack necessity varies under different settings.}
    \label{fig:BF_need_attack}
    \vspace{-0.5mm}
\end{figure}

\begin{figure}[t]
    \centering
    \begin{subfigure}[b]{0.23\textwidth}
        \centering
        \includegraphics[width=\textwidth]{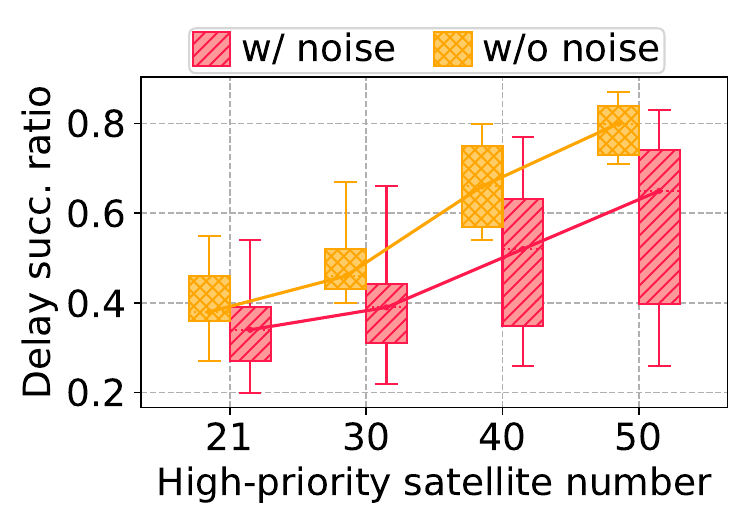}
        \label{fig:delay_success_ratio_vary_skysat}
    \end{subfigure}
    \begin{subfigure}[b]{0.23\textwidth}
        \centering
        \includegraphics[width=\textwidth]{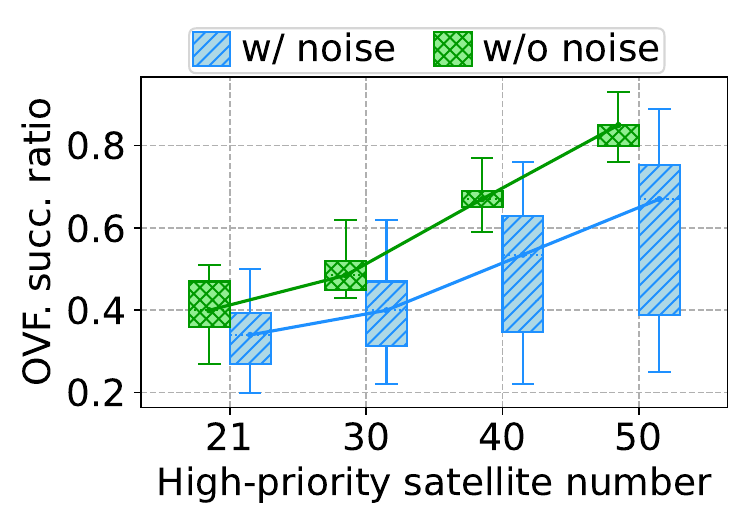}
        \label{fig:BF_success_ratio_vary_skysat}
    \end{subfigure}
    \vspace{-2.5mm}
    \caption{Attack success ratios vs. number of high-priority satellites.}
    \label{fig:success_ratio_vs_skysat}
    \vspace{-0.5mm}
\end{figure}

\begin{figure}[!ht]
    \centering
    \begin{subfigure}[b]{0.23\textwidth}
        \centering
        \includegraphics[width=\textwidth]{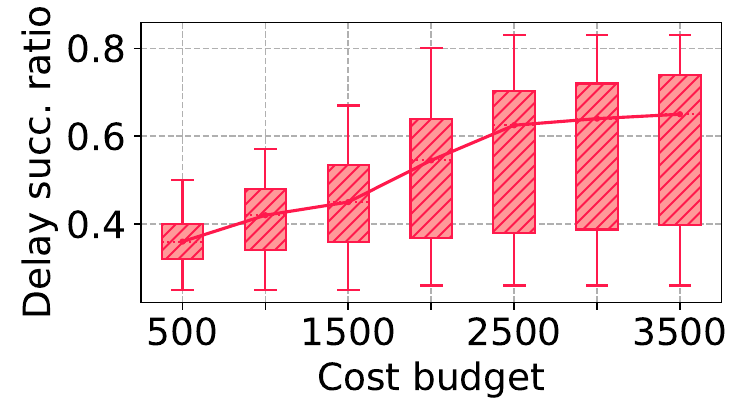}
        \label{fig:delay_success_ratio_vary_cost_budget}
    \end{subfigure}
    \begin{subfigure}[b]{0.23\textwidth}
        \centering
        \includegraphics[width=\textwidth]{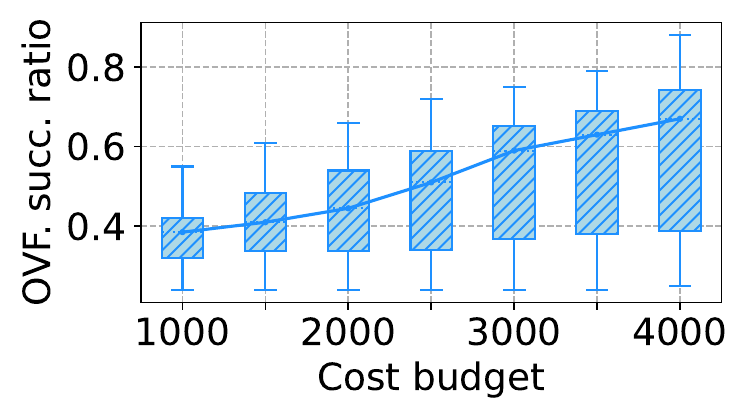}
        \label{fig:BF_success_ratio_vary_cost_budget}
    \end{subfigure}
    \vspace{-2.5mm}
    \caption{Attack success ratios vs. cost budget.}
    \label{fig:success_ratio_vs_cost_budget}
    \vspace{-0.5mm}
\end{figure}

\begin{figure}[t]
    \centering
    \begin{subfigure}[b]{0.23\textwidth}
        \centering
        \includegraphics[width=\textwidth]{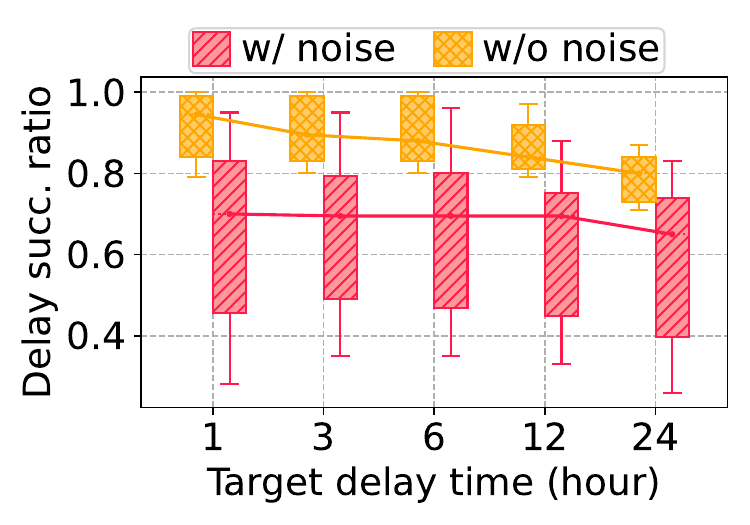}
        \label{fig:delay_success_ratio_vary_target_duration}
    \end{subfigure}
    \vspace{-2.5mm}
    \caption{Attack success ratio vs. target duration.}
    \label{fig:success_ratio_vs_target_duration}
    \vspace{-0.5mm}
\end{figure}

\begin{figure}[!ht]
    \centering
    \begin{subfigure}[b]{0.23\textwidth}
        \centering
        \includegraphics[width=\textwidth]{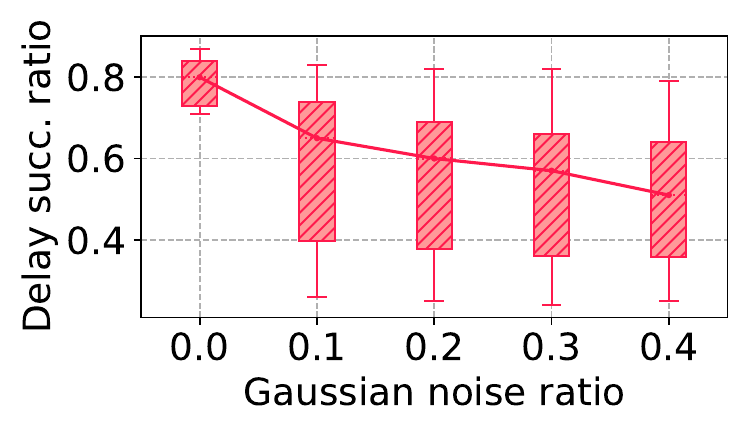}
        \label{fig:delay_success_ratio_vary_noise_combination}
    \end{subfigure}
    \begin{subfigure}[b]{0.23\textwidth}
        \centering
        \includegraphics[width=\textwidth]{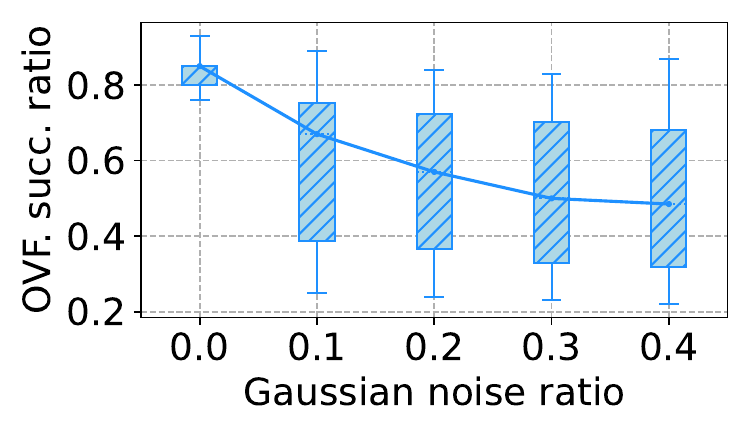}
        \label{fig:BF_success_ratio_vary_noise_combination}
    \end{subfigure}
    \vspace{-2.5mm}
    \caption{Attack success ratios vs. noise ratio.}
    \label{fig:success_ratio_vs_noise_combination}
    \vspace{-0.5mm}
\end{figure}

\begin{figure}[!ht]
    \centering
    \begin{subfigure}[b]{0.23\textwidth}
        \centering
        \includegraphics[width=\textwidth]{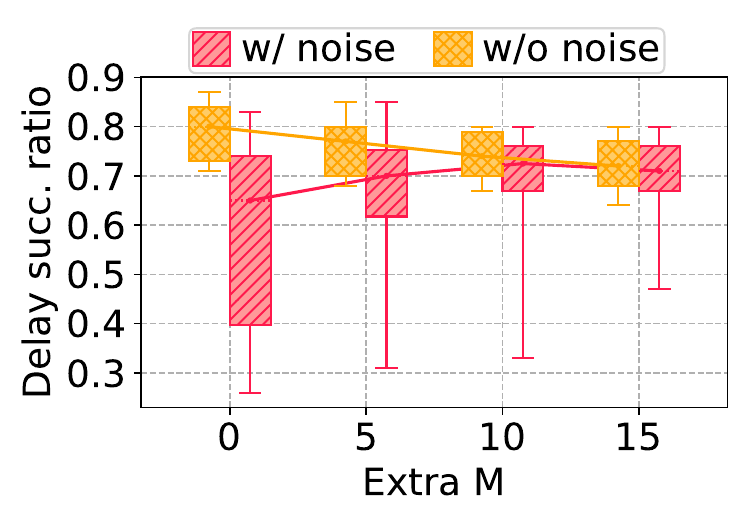}
        \label{fig:delay_success_ratio_vary_M}
    \end{subfigure}
    \begin{subfigure}[b]{0.23\textwidth}
        \centering
        \includegraphics[width=\textwidth]{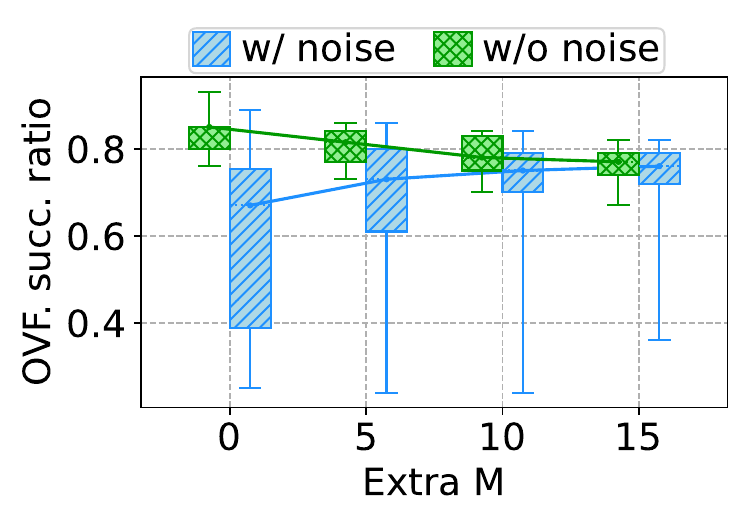}
        \label{fig:BF_success_ratio_vary_M}
    \end{subfigure}
    \vspace{-2.5mm}
    \caption{Attack success ratios vs. extra M.}
    \label{fig:success_ratio_vs_M}
    \vspace{-0.5mm}
\end{figure}

\vspace{-1mm}
\subsection{Evaluation Results}
\label{Ev_2}
\vspace{-1mm}
\noindent We used \emph{Delay succ.\ ratio} and \emph{OVF.\ succ.\ ratio} to denote the data delay attack success ratio and data overflow attack success ratio, which were calculated by the number of successful attacks divided by the total number of attacks attempted.

\noindent \textbf{Vary image size.}
We evaluated the success ratio of the data delay attack and data overflow attack under different image sizes with and without noise, as illustrated in Fig.~\ref{fig:success_ratio_vs_image_size}.
The box in this and the following figures extended from the first quartile to the third quartile of the success ratio, with a line at the median~\cite{box-plot}. The whiskers extended from the box to the maximum and minimum of the success ratio. 
\purple{We connected the median of each box to show the trend as the x-axis changes.}

\purple{Overall, attacks with noise had a lower success ratio than those without, as the attack strategy was generated by the attacker for a noise-free environment.}
For the data delay attack, the success ratio increased as the image size increased. 
\purple{This was because larger images required more time to downlink, giving the attacker more opportunities to block the downlink slots of the target satellite.}
The slight decrease in success ratio observed at the image size of 300MB was attributed to the increased likelihood of data overflow occurrences at this particular image size. Consequently, the available attackable time slots were reduced, making it more challenging to identify an attack strategy capable of delaying the target data for 24 hours.
\purple{The data overflow attack success ratio had a similar trend, with the success ratio increasing as the image size increased.}

\noindent \textbf{Vary data rate.}
\purple{We evaluated success ratio of the two attacks under different data rates, as shown in Fig.~\ref{fig:success_ratio_vs_data_rate}.}
For the data delay attack, the success ratio with noise decreased as the data rate increased. 
This was because a higher data rate allowed the target satellite to downlink images faster, reducing the attacker's chances of blocking the downlink slots.
For the data delay attack without noise, although the success ratio remained relatively stable at 0.8, its trend varied with changes in data rate. Specifically, it decreased from 80Mbps to 160Mbps, then increased from 160Mbps to 320Mbps. 
This pattern arose because at a data rate of 160Mbps, data overflow incidents were more likely to occur, leading to a reduction in available attackable windows and consequently a decrease in success ratio. 
However, as the data rate increased, a larger volume of data was transferred to the ground, making data overflow less likely. This resulted in an increase in available attackable time slots, leading to a higher success ratio.
\purple{The data overflow attack with and without noise showed a similar trend. }

We also analyzed the trend of data overflow success ratio with changes in image size and data rate, both with and without the need for attack (natural data overflow), as shown in Fig.~\ref{fig:BF_need_attack}. 
As shown in the left subfigure, as the image size increased, the overall data overflow success ratio increased, along with a significant increase in natural data overflow success ratio. 
This means that natural data overflow contributed a significant portion to the all data overflow success ratio. 
This was because as the image size increased and the downlink data rate remained unchanged, more data dropped directly due to natural data overflow, without the need for an attack.
Conversely, in the right subfigure, with the increase in data rate, there was an increase in the amount of downlink data per unit time, leading to a decrease in data overflow instances that could occur without the need for an attack. 
As a result, the majority of data overflow successes required the attacker to generate a corresponding attack strategy.

\noindent \textbf{Vary number of high-priority satellites.}
\purple{We evaluated success ratio of the two attacks with varying numbers of high-priority satellites, as shown in Fig.~\ref{fig:success_ratio_vs_skysat}.}
\purple{The number of high-priority satellites significantly impacted the success ratio of both attacks.}
\purple{More high-priority satellites provided more attackable time slots, increasing the success ratio.}

\noindent \textbf{Vary cost budget.}
\purple{We evaluated success ratio of the two attacks under different cost budgets, as shown in Fig.~\ref{fig:success_ratio_vs_cost_budget}.}
The results showed that for an attacker with a limited budget, a higher cost budget allowed the attacker to launch more high-priority tasks, thereby increasing the success ratio of the attack.
The data overflow attack success ratio had a similar trend.

\noindent \textbf{Vary target duration.}
We evaluated success ratio of the data delay attack under different target durations, as shown in Fig.~\ref{fig:success_ratio_vs_target_duration}.
The results indicated that for different target delay times, if the attacker aimed to delay the target data for a longer duration, more attackable time slots were necessary to intercept. However, as the duration increased, more unattackable time slots emerged, potentially allowing the target data to transfer out. Consequently, the effective attackable time slots capable of causing continued delay to the target data decreased, leading to a lower success ratio.

\noindent \textbf{Vary Gaussian noise ratio.}
\purple{We evaluated the impact of Gaussian noise on the success ratio of the two attacks by varying the standard deviation ratio from 0.1 to 0.4.} 
The results showed that as the standard deviation ratio increased, the success ratio of the data delay attack and data overflow attack decreased. This was because the noise introduced uncertainty into the environment, making it more challenging for the attacker to identify the optimal attack strategy, thereby reducing the success ratio of the attack.

\noindent \textbf{Vary extra attack target.}
As discussed in \S\ref{sec:practical-consideration}, there are several ways to help the attacker improve the success ratio of the attack in the scenario where the attacker has imprecise knowledge of the initial queue size, image size in the queue and the data rate varies.
We evaluated one of the strategies which is to attack extra $M$ images around the target data to increase the success ratio of the attack. 
As depicted in Fig.~\ref{fig:success_ratio_vs_M}, without noise, both the success ratios of the data delay attack and data overflow attack decreased because the algorithms needed to generate attack sets targeting more images. This increased demand on the attack strategy led to a lower likelihood of finding suitable attack sets.
With noise, as the number of extra attack targets increased, the success ratio of the data delay attack and data overflow attack increased. This was because the attacker generated attack sets that had redundant attack windows to defend against the noise, thereby increasing the success ratio of the attack.



\section{Countermeasures}
\label{sec:discussion}
\noindent In this section, we discuss several potential countermeasures to the proposed data delay and overflow attacks.

\noindent \textbf{Dynamic priority assignment.} 
\blue{When scheduling high-priority tasks, consideration should be given to the potential impact on future low-priority tasks. Instead of consistently prioritizing high-priority satellites for data transmission, priority can be dynamically adjusted based on various factors, such as the waiting time of low-priority satellites.}

\noindent
\textbf{Introducing unpredictable queue dynamics.} 
Our attacks rely on the adversary's ability to predict queue dynamics of the target satellite. Introducing unpredictable factors will make launching a successful attack much more difficult. 
\blue{Examples include data-specific defenses, such as transferring only the most important images, or assigning higher priority to different portions of the image at different time slots. Data-agnostic defenses might involve using random scheduling policies instead of FIFO (or other deterministic ones), random collaborative data collection across satellites, or random communication scheduling between low- and high-priority constellations.}

\noindent \textbf{Anomaly Detection.} Our algorithms may generate specific (malicious) task scheduling patterns that can be detected by an anomaly detector\blue{, such as by continuously monitoring the average delay or data overflow rate of a satellite}. 

\noindent \textbf{Hardware expansion.} Increasing the onboard storage capacity, deploying more ground stations, enhancing the downlink bandwidth, deploying inter-satellite links, and utilizing spectrum sharing \cite{li2024game} can all enhance the system's robustness.



\section{Conclusion}
\label{sec:conclusion}
\noindent In this paper, we explore a new attack surface that exploits the competition for limited downlink resources among different types of EO satellites. 
We identify two distinct attacks unique to EO constellations: the data delay attack and the data overflow attack. 
In addition, we propose two algorithms aimed at conducting these attacks by finding the minimum-cost attack strategy for data-delay attacks and identifying feasible strategies for data overflow attacks.
We conducted trace-driven simulations using real-world satellite image and orbit data to evaluate attack success ratios under realistic satellite communication settings, demonstrating the effectiveness of \purple{the proposed attacks}.
\dgreen{We hope this work will inspire future research on the new attack surface and the deployment of pro-active countermeasures by constellation operators.}



\section*{Acknowledgment}
\vspace{-1mm}
\noindent
We thank our shepherd Prof.\ Andrei Gurtov and the anonymous reviewers for their valuable comments.
The research of Xiaojian Wang and Ruozhou Yu was sponsored in part by NSF grants 2045539 and 2433966.
The research of Guoliang Xue was sponsored in part by the Army Research Laboratory and was accomplished under Cooperative Agreement Number W911NF-23-2-0225. The views and conclusions contained in this document are those of the authors and should not be interpreted as representing the official policies, either expressed or implied, of the Army Research Laboratory or the U.S. Government. The U.S. Government is authorized to reproduce and distribute reprints for Government purposes notwithstanding any copyright notation herein.
This work utilized data made available through the NASA CSDA Program.



\appendix
\label{sec:appendix}
\red{
\noindent
Here, we provide full proofs for Lemma~\ref{lemma:buffer-overflow-useless} and Theorems~\ref{theorem:target-image-delay-attack-optimal}--\ref{theorem:buffer-overflow-attack-algorithm}.

\begin{center}
\vspace{0.25em}
\textbf{Proof of~Lemma \ref{lemma:buffer-overflow-useless}}
\vspace{0.25em}
\end{center}

\begin{proof}
Given an attack strategy $\mathcal{Y}_{s^*}$, let $t_{\sf ovf}$ be an arbitrary slot in $[t_0, t_{lb}(\tau, \mathcal{Y}_{s*})]$ at which the queue is full under $\mathcal{Y}_{s^*}$, \emph{i.e.}, $Q_{s^*}(t_{\sf ovf}, \mathcal{Y}_{s^*}) = c_{s^*}$.
Further let $t'_{\sf ovf} \in [t_0, t_{\sf ovf})$ be the last slot before $t_{\sf ovf}$ satisfying the same condition as $t_{\sf ovf}$ (when queue is full); if no such slot exists then let $t'_{\sf ovf} = t_0-1$.

To prove Lemma~\ref{lemma:buffer-overflow-useless}, we show that adding any attackable time slot $t' \in \mathcal{A}_{s^*} \cap (t'_{\sf ovf}, t_{\sf ovf}]$ to attack strategy $\mathcal{Y}_{s^*}$ will result in the same queue evolution in both $Q_{s^*}(t, \mathcal{Y}_{s^*} \cup \{ t' \})$ and $Q_{s^*}(\tau, t, \mathcal{Y}_{s^*} \cup \{ t' \})$ as before adding $t'$, for any $t \ge t_{\sf ovf}$.

First, consider any slot $t \in [t', t_{\sf ovf}]$.
By definition, when $t = t'$, the full queue $Q_{s^*}(t, \mathcal{Y}_{s^*} \cup \{ t' \}) = \min \{ c_{s_*}, Q_{s^*}(t, \mathcal{Y}_{s^*}) + \alpha^\tau_{s^*}(t') \}$, where $\alpha^\tau_{s^*}(t')$ is the additional residual data amount due to transmission slot $t'$ being attacked.
Similarly, the subqueue before target data $\tau$ will become $Q_{s^*}(\tau, t, \mathcal{Y}_{s^*} \cup \{ t' \}) = Q_{s^*}(\tau, t, \mathcal{Y}_{s^*}) + \alpha^\tau_{s^*}(t')$.

If $t' = t_{\sf ovf}$, we have a full queue $Q_{s^*}(t', \mathcal{Y}_{s^*}) = c_{s^*}$, and so $Q_{s^*}(t', \mathcal{Y}_{s^*} \cup \{ t' \}) = \min\{ c_{s^*}, Q_{s^*}(t', \mathcal{Y}_{s^*}) + \alpha^\tau_{s^*}(t') \} = c_{s^*}$.
In this case, the downlink data amount decreases to $O_{s^*}(t', \mathcal{Y}_{s^*} \cup \{ t' \}) = O_{s^*}(t', \mathcal{Y}_{s^*}) - \alpha^\tau_{s^*}(t')$ due to the attack, and the dropped amount increases to $D_{s^*}(t', \mathcal{Y}_{s^*} \cup \{ t' \}) = D_{s^*}(t', \mathcal{Y}_{s^*}) + \alpha^\tau_{s^*}(t')$ due to additional queue overflow.
Hence, the subqueue before target data $\tau$ also remains the same: $Q_{s^*}(\tau, t', \mathcal{Y}_{s^*} \cup \{ t' \}) = Q_{s^*}(\tau, t', \mathcal{Y}_{s^*})$.
Since both $Q_{s^*}(t', \mathcal{Y}_{s^*})$ and $Q_{s^*}(\tau, t, \mathcal{Y}_{s^*} \cup \{ t' \})$ remains unchanged at $t' = t_{\sf ovf}$, the queue evolution after $t_{\sf ovf}$ is unaffected.

If $t' < t_{\sf ovf}$, the remaining queue at time $t'$ is increased by $\alpha^\tau_{s^*}(t')$, that is, $Q_{s^*}(t', \mathcal{Y}_{s^*} \cup \{ t' \}) = Q_{s^*}(t', \mathcal{Y}_{s^*}) + \alpha^\tau_{s^*}(t')$.
By induction, we have $Q_{s^*}(t, \mathcal{Y}_{s^*} \cup \{ t' \}) = Q_{s^*}(t, \mathcal{Y}_{s^*}) + \alpha^\tau_{s^*}(t')$ for each time slot $t \in [t', t_{\sf ovf})$.
At time $t = t_{\sf ovf}$, since $Q_{s^*}(t, \mathcal{Y}_{s^*}) = c_{\s^*}$, this leads to the same derivation as above where $Q_{s^*}(t, \mathcal{Y}_{s^*} \cup \{ t' \}) = \min\{ c_{s^*}, Q_{s^*}(t, \mathcal{Y}_{s^*}) + \alpha^\tau_{s^*}(t') \} = c_{s^*}$.
Similarly, we have $Q_{s^*}(\tau, t, \mathcal{Y}_{s^*} \cup \{ t' \}) = Q_{s^*}(\tau, t, \mathcal{Y}_{s^*}) + \alpha^\tau_{s^*}(t')$ for each time slot $t \in [t', t_{\sf ovf})$ due to the downlink amount $O_{s^*}(t', \mathcal{Y}_{s^*} \cup \{ t' \})$ being reduced by $\alpha^\tau_{s^*}(t')$ at time $t'$.
At time $t = t_{\sf ovf}$, the dropped amount $D_{s^*}(t, \mathcal{Y}_{s^*} \cup \{ t' \})$ increased by $\alpha^\tau_{s^*}(t')$ due to queue overflow, and hence the subqueue $Q_{s^*}(\tau, t, \mathcal{Y}_{s^*} \cup \{ t' \})$ still remains the same.
Hence, the queue evolution after $t_{\sf ovf}$ remains unaffected.

Summarizing the above, since the queue evolution of both the full queue and the subqueue before $\tau$ remains unchanged after $t_{\sf ovf}$, the expected evacuation time of the target data $\tau$ remains unchanged.
By definition of $t_{lb}(\tau, \mathcal{Y}_{s*})$, this means attacking any single slot before $t_{lb}(\tau, \mathcal{Y}_{s*})$ results in no attack strength.
The conclusion extends to an arbitrary set of attack slots before $t_{lb}(\tau, \mathcal{Y}_{s*})$ by induction over the set, thus proving Lemma~\ref{lemma:buffer-overflow-useless}.
\end{proof}

\newcommand{\oldY}{{\mathcal{Y}_{s^*}^{t}}}
\newcommand{\newY}{{\mathcal{\overline Y}_{s^*}^{\textsf{next}(t)}}}

\begin{center}
\vspace{0.25em}
\textbf{Proof of Theorem~\ref{theorem:target-image-delay-attack-optimal}}
\vspace{0.25em}
\end{center}

\begin{proof}
We prove Theorem~\ref{theorem:target-image-delay-attack-optimal} by induction.

Let $\textsf{next}(t) = \min_{t'} \{t'| t' \in \mathcal{X}_{s^*} \text{ and } t' > {t}\}$ be the next transmissible time slot after time $t$. 
Let $\delta_{s^*}^t$ be the number of transmissible time slots between the original (non-attacked) evacuation time $t_e(\tau, \emptyset)$ and the delay target $t$.
Assume the data delay attack problem is feasible for a target delay $t^*(\tau) = t$, we define a \textbf{minimal attack strategy (MAS)} to be a feasible attack strategy $\mathcal{Y}_{s^*}$ for target delay $t$, such that removing any slot from $\mathcal{Y}_{s^*}$ results in it no longer being feasible for $t$.
By definition, a MAS has the following properties:
\begin{enumerate}
    \item A MAS for $t$ must have size equal to $\delta_{s^*}^t$;
    \item A MAS for $\textsf{next}(t)$ has exactly one more time slot than a MAS for $t$;
    \item A minimum-cost attack strategy for $t$ must be a MAS (but not vice versa).
\end{enumerate}
The third property above is true because, if a minimum-cost attack strategy $\mathcal{Y}_{s^*}$ is not a MAS, then there exists at least one time slot in $\mathcal{Y}_{s^*}$, such that removing it still results in a feasible attack strategy with less cost.

Based on the above, we prove the following statement by induction: \emph{in every iteration $i$ of Algorithm~\ref{al:target_image_delay}, it finds a minimum-cost attack strategy with target delay equal to the $i$-th transmissible time slot after $t_e(\tau, \emptyset)$.}

\textbf{Base step:}
Before the first iteration ($i = 0$), the statement is true, since $\mathcal{Y}_{s^*} = \emptyset$ is a minimum-cost attack strategy for target delay equal to the original evacuation time $t_e(\tau, \emptyset)$.

\textbf{Induction step:}
Let $t$ be the $i$-th transmissible time slot after $t_e(\tau, \emptyset)$.
Assume target delay $t$ is feasible, and our algorithm has found a minimum-cost attack strategy ${\oldY}$ for $t$ at iteration $i$.
Further assume target delay $\textsf{next}(t)$ is also feasible.
We will prove that our algorithm finds a minimum-cost attack strategy for $\textsf{next}(t)$ at iteration $i+1$.

Our induction proof is divided into two parts: 
(1) Algorithm~\ref{al:target_image_delay} finds a \emph{minimal} (and thus feasible) attack strategy for $\textsf{next}(t)$ at iteration $i+1$.
(2) The attack strategy has the minimum cost among all feasible strategies for $\textsf{next}(t)$.

\vspace{0.25em}
\emph{\underline{Part (1): Feasibility and Minimality.}}
Let ${\newY}$ be an arbitrary MAS (which is feasible) for target delay $\textsf{next}(t)$.
We now describe a step-by-step process for transforming ${\newY}$ into an MAS for $\textsf{next}(t)$ that has the exact shape of ${\oldY} \cup \{ \bar t \}$, \emph{i.e.}, the solution we found in iteration $i$ plus an extra element $\bar t$.
{Each step replaces one element $t_2 \in {\newY}$ but not in ${\oldY}$ with one $t_1 \in {\oldY} \setminus {\newY}$, until when ${\oldY} \subset {\newY}$.}
During the transformation, we prove that after each step: \\
(a) the attack strategy is still \emph{feasible}; \\
(b) the attack strategy is still \emph{minimal} (having size $\delta_{s^*}^t+1$); \\
(c) the intersection between ${\newY}$ and ${\oldY}$ increases by $1$.

Consider $t_1 \in {\oldY} \setminus {\newY}$, and let $\mathcal{Y}' = {\newY} \cup \{ t_1 \}$.
Since ${\oldY} \not\subseteq {\newY}$ until termination, such $t_1$ must exist.
\textbf{Two cases} may arise: either $\mathcal{Y}'$ results in the evacuation time of the target data until $\textsf{next}(\textsf{next}(t))$---resulting in an extra delay of the target data---or the evacuation remains the same at $\textsf{next}(t)$.
Both cases are discussed below:

\vspace{0.25em}
\emph{$\vartriangleright$ Case 1: adding $t_1$ results in additional delay.}
In this case, we find any $t_2 \in {\newY} \setminus {\oldY}$, and remove $t_2$ from ${\newY}$.
Such $t_2$ must exist since $|{\newY}| > |{\oldY}|$.
After this:

(a) Since before removing $t_2$, the strategy ${\newY}$ delays target data until $\textsf{next}(\textsf{next}(t))$, then after removing $t_2$, the expected evacuation time is still at least $\textsf{next}(t)$, proving \emph{feasibility} of the new ${\newY}$ for target delay $\textsf{next}(t)$.

(b) The size of ${\newY}$ remains unchanged before adding $t_1$ versus after removing $t_2$, and hence ${\newY}$ is still \emph{minimal}.

(c) Replacing $t_2$ with $t_1$ results in one more element that ${\newY}$ and ${\oldY}$ have in common.

\vspace{0.25em}
\emph{$\vartriangleright$ Case 2: adding $t_1$ results in \textbf{no} additional delay.}
This means a buffer overflow has happened after $t_1$, resulting in zero attack strength of $t_1$ according to Lemma~\ref{lemma:buffer-overflow-useless}.
{Let $t_{lb}^{t_1} < \textsf{next}(t)$ be the earliest time slot after $t_1$ that has a full queue before adding $t_1$ to ${\newY}$.}
Assume we can find $t_2 \in {\newY}$ such that $t_2 \le t_{lb}^{t_1}$, then removing $t_2$ from ${\newY}$ will result in the same queue evolution after $t_{lb}^{t_1}$, following the same proof logic of Lemma~\ref{lemma:buffer-overflow-useless}.
In this case, we have: (a) the resulting ${\newY}$ is still \emph{feasible} since removing $t_2$ does not result in reduction of evacuation time; (b) ${\newY}$ is still \emph{minimal}; (c) ${\newY}$ has one more common element with ${\oldY}$.

We now show that there must exist such a $t_2 \in {\newY}$ and $t_2 \le t_{lb}^{t_1}$.
Assume such a $t_2$ does not exist.
This means that even without any attack before $t_{lb}^{t_1}$, the queue is still full at $t_{lb}^{t_1}$.
In this case, adding $t_1$ to any attack strategy---including ${\oldY} \setminus \{ t_1 \}$---will result in zero attack strength by Lemma~\ref{lemma:buffer-overflow-useless}.
This contradicts ${\oldY}$ being \emph{minimal} in the induction hypothesis.
Therefore, there must exist at least one element $t_2 \le t_{lb}^{t_1}$ that is in the attack strategy ${\newY}$.

To summarize, in whichever case above, one can always find a $t_2 \in {\newY} \setminus {\oldY}$, and replace it with a $t_1 \in {\oldY} \setminus {\newY}$, which preserves both feasibility and minimality of ${\newY}$.
After up to $\delta^t_{s^*}$ steps, the attack strategy ${\newY}$ becomes a MAS in the shape of ${\oldY} \cup \{ \bar t \}$.

\textbf{\underline{Remark:}} Note that the above transformation is valid not just between ${\newY}$ and ${\oldY}$, but also any pair of MASs.
This fact will be used in Part (2) of this proof, as well as the next proof for Theorem~\ref{theorem:buffer-overflow-attack-algorithm} that is based on a similar idea.

Finally, to complete the proof of Part (1), as long as a feasible solution ${\oldY} \cup \{ \bar t \}$ exists for target delay $\textsf{next}(t)$, Algorithm~\ref{al:target_image_delay} can always find such a $\bar t$ in Line~\ref{line:find_min_cost_time_slot}.
To show this, note that if $\bar t$ exists before $t_{lb}(\tau, {\oldY})$ in the algorithm, then by Lemma~\ref{lemma:buffer-overflow-useless} it will have zero attack strength.
In this case, adding $\bar t$ to ${\oldY}$ leads to a non-minimal attack strategy, violating our proof above that ${\oldY} \cup \{ \bar t \}$ is a MAS.
Hence $\bar t$ (which earlier proof shows to exist) must be in the range $\hat T_{\tau}$ that the algorithm searches within, proving $\hat T_{\tau}$ to be non-empty.
By finding a slot with non-zero attack strength, the algorithm finds one such $\bar t$, leading to a MAS feasible to $\textsf{next}(t)$ in iteration $i+1$.

\newcommand{\newnewY}{\mathcal{Y}_{s^*}}
\newcommand{\newminY}{\mathcal{\overline Y}_{s^*}}

\emph{\underline{Part (2): Optimality.}}
To prove optimality, we show during the induction step that, if iteration $i$ generates ${\oldY}$ that is a minimum-cost MAS for target delay $t$, then iteration $i+1$ generates a ${\mathcal{Y}_{s^*}^{\textsf{next}(t)}} = {\oldY} \cup \{ \bar t \}$ that is a minimum-cost MAS for target delay $\textsf{next}(t)$, by adding $\bar t$ in iteration $i$.
Below, we drop the superscript $\textsf{next}(t)$ when referring to a solution for target delay $\textsf{next}(t)$ for simplicity.

Let ${\newminY}$ be a minimum-cost MAS for $\textsf{next}(t)$.
Define two sets of time slots: $\mathcal{Y}_1 = {\newnewY} \setminus {\newminY}$, and $\mathcal{Y}_2 = {\newminY} \setminus {\newnewY}$.
Let $\Upsilon_{\mathcal{Y}}$ be the cost of an arbitrary strategy $\mathcal{Y}$.
Proving the algorithm's output ${\newnewY}$ is minimum-cost means showing that the cost of ${\newnewY}$ is at most the cost of the minimum-cost ${\newminY}$, in other words, $\Upsilon_{\mathcal{Y}_1} \le \Upsilon_{\mathcal{Y}_2}$.

Assume by contradiction $\Upsilon_{\mathcal{Y}_1} \!\!>\!\! \Upsilon_{\mathcal{Y}_2}$.
Consider \textbf{two cases}: if $\mathcal{Y}_1$ is a subset of $\mathcal{Y}_{s^*}^t$ (the iteration-$i$ solution), or not.

\vspace{0.25em}
\emph{$\vartriangleright$ Case 1: $\mathcal{Y}_1 \subseteq \mathcal{Y}_{s^*}^t$.}
Following the same procedure in Part (1) of the proof, we can step-by-step replace all elements in $\mathcal{Y}_1$ with all elements in $\mathcal{Y}_2$ from ${\oldY}$.
This results in a feasible solution to target delay $t$ with less cost than ${\oldY}$, contradicting that ${\oldY}$ is minimum-cost in the induction hypothesis.

\vspace{0.25em}
\emph{$\vartriangleright$ Case 2: $\mathcal{Y}_1 = \mathcal{Y}_1' \cup \{ \bar t \}$ where $\mathcal{Y}_1' \subseteq \mathcal{Y}_{s^*}^t$.}
In this case, the lower cost of $\mathcal{Y}_1$ may be due to either $\mathcal{Y}_1'$ or $\bar t$.
Based on the replacement procedure in Part (1), if we can find a time slot $\bar t_2 \in \mathcal{Y}_2$ whose cost is lower than $\bar t$ and who can replace $\bar t$ without affecting feasibility, this time slot must be in the range $\hat T_\tau$ during iteration $i+1$, which contradicts the fact that our algorithm picks the minimum-cost attackable slot in $\hat T_\tau$.

Assume such a slot $\bar t_2$ cannot be found in $\mathcal{Y}_2$.
It then means there is a subset $\mathcal{Y}_2' \subset \mathcal{Y}_2$, such that: (i) $\Upsilon_{\mathcal{Y}'_2} < \Upsilon_{\mathcal{Y}'_1}$, and (ii) replacing all elements in $\Upsilon_{\mathcal{Y}'_1}$ with all elements in $\Upsilon_{\mathcal{Y}'_2}$ from ${\oldY}$ results in a feasible solution for target delay $t$ with less cost.
This again contradicts with ${\oldY}$ being minimum-cost.

Summing up the above cases, by induction, the solution ${\newnewY}$ output by the algorithm in iteration-$(i+1)$ is a minimum-cost attack strategy for target delay $\textsf{next}(t)$.
\end{proof}

\begin{center}
\textbf{Proof of Theorem~\ref{theorem:buffer-overflow-attack-algorithm}}
\end{center}

\begin{proof}
The proof of Theorem~\ref{theorem:buffer-overflow-attack-algorithm} follows the same thread as Theorem~\ref{theorem:target-image-delay-attack-optimal}, based on step-by-step transformation from any feasible solution to a solution that our algorithm can find.

Similar to the previous proof, we define two concepts:
A \textbf{minimal attack strategy (MAS)} $\mathcal{Y}_{s^*}$ is a feasible attack strategy of the data overflow attack, removing any element from which results in infeasibility of the attack.
An \textbf{earliest attack strategy (EAS)} is a feasible attack strategy that results in the earliest time slot $t_d$ at which the target data $\tau$ is dropped.
Both types of strategy must exist if data overflow attack is feasible for the target data.
Our proof specifically focuses on strategies that are \emph{both a MAS and an EAS}: a \textbf{minimal earliest attack strategy (MEAS)}.

\newcommand{\oldYhere}{\mathcal{Y}_{s^*}}

We now show that all MEASs for dropping the target data have the same size.
If the target data is dropped without any attack, then the only MEAS is an empty strategy $\emptyset$.
Assuming target data is not dropped when no attack happens, we have $t_d > t_e(\tau, \emptyset)$, meaning the attack is feasible due to the target data's downlink time being delayed until at least $t_d$.
Any MEAS of overflow attack must be a MAS of the data delay attack that delays the target data until the next transmissible time slot---denoted by $\textsf{tx}(t_d)$---on or after $t_d$.
To see this, note that if the evacuation time under a strategy is before $t_d$, the target data would be downlinked and the attack fails.
If the evacuation time is after $\textsf{tx}(t_d)$, then because target data $\tau$ would not be at the queue head at time $t_d$, it would not be dropped due to buffer overflow at $t_d$, making such a strategy either infeasible or not an EAS.
Then, by the same argument as in the last proof, all MEASs have the same size $\delta_{s^*}^{\textsf{tx}(t_d)}$.

Given the above, a MEAS for the data overflow attack is always a MAS for the data delay attack for a fixed target delay $\textsf{tx}(t_d)$ (assuming, with a slight modification of definition, that the queue does not overflow between $t_d$ and $\textsf{tx}(t_d)$ and keeps the target data on-board instead).
By the same induction argument as in Part (1) of the last proof, we can immediately prove that any MAS to such a data delay attack can be transformed to a (data delay) MAS that our algorithm can find, starting from an empty MAS for the no attack scenario.
By a similar argument as in Part (2) of the last proof, we can also prove that our algorithm finds a MEAS of the data overflow attack.

Summing up the above, Algorithm~\ref{alg:buffer_overflow} always finds a feasible attack strategy to the data overflow attack when one exists, under the condition specified in Theorem~\ref{theorem:buffer-overflow-attack-algorithm}.
\end{proof}

}


\bibliographystyle{IEEEtran}



\end{document}